\documentclass[10pt,preprint]{aastex}
\usepackage{epsfig}

\newcommand{\myemail}{takahasi@phyas.aichi-edu.ac.jp}

\shorttitle{TRANS-MAGNETOSONIC ACCRETION IN A BLACK HOLE MAGNETOSPHERE}
\shortauthors{ M. Takahashi }

\begin{document}

\title{TRANS-MAGNETOSONIC ACCRETION IN A BLACK HOLE MAGNETOSPHERE}

\author{ MASAAKI TAKAHASHI }
\affil{  Department of Physics and Astronomy, 
         Aichi University of Education, \\
         Kariya 448-8542, Japan}
\email{\myemail}

\begin{abstract} 
We present the critical conditions for hot trans-fast magnetohydrodynamical 
(MHD) flows in a stationary and axisymmetric black-hole magnetosphere. 
To accrete onto the black hole, the MHD flow injected from a plasma source 
with low velocity must pass through the fast magnetosonic point after 
passing through the ``inner'' or ``outer'' Alfv\'en point. We find that 
a trans-fast MHD accretion solution related to the inner Alfv\'en point 
is invalid when the hydrodynamical effects on the MHD flow dominate at 
the magnetosonic point, while the other accretion solution related to 
the outer Alfv\'en point is invalid when the total angular momentum of 
the MHD flow is seriously large. 
When both regimes of the accretion solutions are valid in the black hole 
magnetosphere, we can expect the transition between the two regimes. 
The variety of these solutions would be important in many highly energetic 
astrophysical situations. 
\end{abstract}


\section{Introduction}

In order to explain the activity of active galactic nuclei (AGNs) and 
compact X-ray sources, we consider a black hole magnetosphere in the 
center of these objects. The magnetosphere is composed of a central 
black hole with surrounding plasmas and a large scale magnetic field. 
The magnetic field is originated from an accretion disk rotating 
around the black hole. The electrodynamics of the black hole 
magnetosphere has been discussed by many authors; force-free
magnetospheres were discussed in \citet{Membrane86} and more 
general magnetospheres in \citet{Punsly01}.

In the black hole magnetosphere, because of the strong gravity of 
the black hole and the rapid rotation of the magnetic field, both 
an ingoing plasma flow (accretion) and an accelerated outgoing plasma 
(wind/jet) should be created.  
The plasma would be provided from the disk surface and its corona. 
When the plasma density in the magnetosphere is somewhat large, 
the plasma inertia effects should be important. In this case, the 
plasma would be nearly neutral and should be treated by the ideal 
magnetohydrodynamic (MHD) approximation \citep{Phinney83}, so the 
plasma streams along a magnetic field line, where the magnetic field 
line could extend from the disk surface to the event horizon or 
a far distant region (\cite{Nitta-TT91}; see also \cite{TT2001}). 
The outgoing flow effectively carries the angular momentum from 
the plasma source, and then the accretion would continue to be 
stationary, releasing its gravitational energy. 
The magnetic field lines connecting the black hole with the disk, 
which are mainly generated by the disk current, may not connect 
directly to the distant region, but via the disk's interior the 
energy and angular momentum of the black hole can be carried to 
the distant region; the energy and angular momentum transport 
inside the disk is not discussed here.

If the plasma density is sufficiently low and the magnetosphere is 
magnetically dominated, one can expect the pair-production region 
along open magnetic flux tubes, which connect the black hole to far 
distant regions directly \citep{beskin97, Punsly01}. Further, we also 
expect the Blandford-Znajek (1977) process, which suggests the 
extraction of energy and angular momentum from the spinning black hole.

In this paper, we assume a stationary and axisymmetric magnetosphere, 
and consider ideal MHD flows along a magnetic field line. The initial 
velocity can be at most less than the slow magnetosonic wave speed. 
To accrete onto the black hole, the ejected inflows from the the plasma 
source must pass through the slow magnetosonic point ({\sf S}), the 
Alfv\'en point ({\sf A}) and the fast magnetosonic point ({\sf F}) 
in this order, as it is well known. At these points, {\sf A}, {\sf F} 
and {\sf S}, the poloidal velocity equals one of the Alfv\'{e}n wave 
and fast and slow magnetosonic wave speeds, respectively. 
In the case of accretion onto a star, because the accreting plasma 
is stopped at the stellar surface, a shock front would be formed 
somewhere on the way to the stellar surface and the accretion becomes 
sub-fast magnetosonic. 
However, for accretion onto a black hole, the flow must be 
super-fast magnetosonic at the event horizon ({\sf H}). If not so, 
the fast magnetosonic wave can extract information from the interior 
of the black hole to the exterior; this fact obviously contradicts 
with the definition of the event horizon. 
In fact, an ideal MHD accretion solution which keeps sub-fast 
magnetosonic has zero poloidal velocity at the event horizon and the 
density of the plasma diverges; the solution is unphysical.

Because the magnetic field lines would rigidly rotate under the ideal 
MHD assumption, there are two light surfaces ({\sf L}) in the black 
hole magnetosphere \citep[hereafter Paper~I]{Znajek77,Takahashi-NTT90}. 
The plasma source must be located between these two surfaces. Further, 
one or two Alfv\'en surfaces lie between the two light surfaces, 
and for accretion there must be a fast-magnetosonic surface 
between the Alfv\'en surface and the event horizon (see Paper~I).
Here, we should note that the physical mechanism to determine the 
angular velocity of the field lines is controversial. A time-dependent 
determination of it has been discussed by \citet{Punsly01}; 
the torsional Alfv\'en wave originated from the plasma source and 
propagated up and down the magnetic flux tube forces to minimize the 
magnetic stresses in the system.

The conditions on the flows at the magnetosonic points and the Alfv\'en 
point restrict the five physical parameters which specify the flow 
(see the following section) if one is to adhere to the ideal MHD 
assumption globally. The fast and slow critical points have X-type 
(physical) or O-type (unphysical) topology for the solution, while the 
Alfv\'{e}n point does not specify its topological feature; hereafter, 
we will call the flow passing through both X-type fast and X-type 
slow magnetosonic points as the ``{\sf SAF}-solution''. When we discuss the 
global features of a solution, it is very important to know the numbers 
of these critical points and the Alfv\'en points.  
In a zero-temperature limit (cold limit), the regularity condition 
for the trans-fast MHD flow was discussed by \citet{Takahashi94}. 
In this case, the Alfv\'en points and the fast magnetosonic points only 
appear in wind and accretion solutions without the slow magnetosonic 
point, because the velocity of the slow magnetosonic wave speed is zero. 
The relativistic hot MHD flow equation has been formulated by 
\citet{Camenzind86a,Camenzind86b,Camenzind87,Camenzind89}; 
see also, Paper~I. In \S~2, we summarize the basic equations for the 
MHD flows and the condition at the Alfv\'en point discussed in Paper~I.

Though we consider the general relativistic plasma flow, the 
significance of the Alfv\'en point and the fast and slow magnetosonic
point conditions is similar to that of a Newtonian wind model by
\citet{Weber-Davis67} and a special relativistic wind model by
\citet{Kennel83}. \citet{Kennel83} classified the outgoing
trans-Alfv\'enic MHD wind solutions into a ``critical'' ($E=E_{\rm F}$) 
solution, ``sub-critical'' ($E<E_{\rm F}$) solutions and
``super-critical'' ($E>E_{\rm F}$) solutions, where $E$ is the conserved
energy of the wind and $E_{\rm F}$ is the energy for the trans-fast 
MHD wind. To reach distant regions, the critical (trans-fast MHD wind) 
solution and the super-critical (sub-fast MHD winds) solutions are 
physical, and the sub-critical solutions are unphysical beyond the 
turnaround point. 
Turning now to the ingoing plasma flow, the topology of the black hole 
accretion solution space also has a similar structure; that is, 
(i) a trans-Alfv\'en MHD ingoing flow with $E = E_{\rm F}$, which 
    means a trans-fast MHD ingoing flow discussed in this paper 
    (critical), 
(ii) trans-Alfv\'en MHD ingoing flows with $E < E_{\rm F}$ 
     (sub-critical) and 
(iii) trans-Alfv\'en MHD ingoing flows with $E > E_{\rm F}$ 
      (super-critical). 
Under the ideal MHD approximation, the physical solution is only the 
critical solution (i); the super-critical solutions (iii) are unphysical 
for the reason mentioned above. 
In addition to these, for accretion onto a black hole, we must consider 
(iv) sub-Alfv\'en (or sub-slow MHD) ingoing flows, although they do not 
     pass through the Alfv\'en point {\sf A}. 
The breakdown of ideal MHD approximation between the horizon and 
the inner light surface is indicated by \citet{Punsly01}, and then 
non-ideal MHD solutions classified into (ii), (iii) and (iv) would 
be realized as accretion solutions onto a black hole (see \S~5).

The main purpose of this paper is to examine the thermal effects on 
an ideal MHD plasma streaming in a black hole magnetosphere 
(see Fig.~\ref{fig:acc}) by studying the critical conditions at 
those magnetosonic points. Now, the slow magnetosonic point appears 
on the MHD flow solutions. The details of critical conditions at the 
fast and slow magnetosonic points are discussed in \S~3. 
We derive the critical conditions at the fast and slow magnetosonic 
points, which are denoted in terms of the location of the fast and 
slow magnetosonic points, the sound velocity at the fast and slow 
magnetosonic points and the locations of the Alfv\'{e}n and light 
surfaces. 
In \S~4, we clarify the thermal effects on the MHD flows, and discuss 
its dependence on the rotation of the black hole magnetosphere and the 
divergence of the cross-section of a magnetic flux-tube along the field 
line. Then, we can find two kinds of trans-fast MHD flow solutions for 
both inflows and outflows: ``hydro-like'' MHD flow and ``magneto-like'' 
MHD flow. The main difference between the two solutions is the behavior 
of the magnetization parameter, which is the ratio of the fluid and 
electromagnetic parts of the total energy of the flow. The hydro-like 
MHD flow solution is a somewhat hydrodynamical solution and, in the weak 
magnetic field limit, this trans-fast magnetosonic flow solution becomes 
a trans-sonic flow solution discussed by \citet{Abramowicz81} and 
\citet{Lu86} in the hydrodynamical case. We also unify hydrodynamic 
flows with hot MHD flows in a common formalism. The hydro-like MHD flow 
solution disappears for a magnetically-dominated magnetosphere. 
On the contrary, the magneto-like MHD flow solution results in the 
magnetically-dominated flow, although that disappears for hotter plasma 
cases. 
In \S~5, we summarize our results.

\placefigure{fig:acc}

\section{Basic Equations and Trans-Alfv\'{e}nic MHD Flows }

We present basic equations of a stationary and axisymmetric 
ideal MHD flow. The flow streams along a magnetic field line in the 
black hole magnetosphere, and accretes onto the black hole or blows 
away to a far distant region. 
To determine the configuration of magnetic field lines and the 
velocity of MHD flows streaming along each magnetic field line, we must 
solve self-consistently what is called the Bernoulli equation along 
magnetic field lines and the magnetic force-balance equation. 
These equations are derived from the equation of motion for relativistic 
MHD plasma 
\begin{equation}
   T^{\mu\nu}_{;\nu} = \left[(\rho+P)u^\mu u^\nu - Pg^{\mu\nu}
       + \frac{1}{4\pi} \left(F^{\mu\delta}F_{\delta}^{\ \nu} +
         \frac{1}{4}g^{\mu\nu}F^2\right)  
         \right]_{;\nu} = 0 \ ,
\end{equation}
the conservation law for particle number $(nu^\mu)_{;\mu}=0$, the 
ideal MHD condition $u^\mu F_{\mu\nu}=0$ and Maxwell's equations. 
Here, $\rho$, $P$ and $n$ are the total energy density, the pressure of 
the plasma and the proper particle number density. The electromagnetic 
field tensor $F^{\mu\nu}$ satisfies Maxwell equations and $u^\mu$ is 
the four-velocity of the plasma. 
The background metric is written by the Boyer-Lindquist coordinates 
with $c=G=1$, 
\begin{eqnarray}
   ds^2 &=& \left( 1-\frac{2mr}{\Sigma} \right) dt^2 
        + \frac{4amr\sin^2\theta}{\Sigma} dt d\phi \\
       & &  - \left( r^2+a^2+\frac{2a^2mr\sin^2\theta}{\Sigma} \right)
                                     \sin^2\theta d\phi^2
            - \frac{\Sigma}{\Delta} dr^2 - \Sigma d\theta^2  \ ,  \nonumber 
\end{eqnarray}
where $\Delta \equiv r^2 -2mr +a^2 $, $\Sigma \equiv r^2 +a^2\cos^2\theta$,
and  $m$ and $a$ denote the mass and angular momentum per unit mass 
of the black hole, respectively.

The flow with the above assumptions streams along a magnetic field line, 
which is expressed by a magnetic stream function $\Psi=\Psi(r,\theta)$ 
with $\Psi=$ constant; $\Psi$ is basically the toroidal component 
of the vector potential. Stationarity, axisymmetry and ideal MHD 
condition require the existence of five constants of motion (e.g., 
Bekenstein \& Oron 1978; Camenzind 1986a). These conserved quantities 
are the total energy $E(\Psi)$, the total angular momentum $L(\Psi)$, 
the angular velocity of the field line $\Omega_F(\Psi)$ and the particle 
flux through a flux tube $\eta(\Psi)$, which are given by
\begin{eqnarray}
        E &=&  \mu u_t - \frac{\Omega_F}{4\pi\eta}  B_\phi \ ,\label{eq:EE} \\
        L &=& -\mu u_\phi - \frac{1}{4\pi\eta}      B_\phi \ ,\label{eq:LL} \\
 \Omega_F &=& -\frac{F_{tr}}{F_{\phi r}} \ = \
                  -\frac{F_{t\theta}}{F_{\phi \theta}}  \ , \\
     \eta &=& \frac{nu_p}{B_p} \ ,
\end{eqnarray}
where $B_\phi \equiv (\Delta/\Sigma)\sin\theta\; F_{\theta r}$ is 
the toroidal component of the magnetic field, 
$B_p$ is the poloidal component of the magnetic field 
seen by a lab-frame observer
\begin{equation}
   B_p^2 \equiv -\frac{1}{\rho_w^2}\left[ g^{rr}(\partial_r\Psi)^2 
          + g^{\theta\theta}(\partial_\theta\Psi)^2 \right]\ ,
\end{equation}
and $\rho_w^2\equiv g_{t\phi}^2-g_{tt}g_{\phi\phi}$. 
The poloidal component $u_p$ of the velocity is defined by 
$u_p^2\equiv -u_Au^A$ ($A=r$, $\theta$), where we set $u_p>0$ $(u^r<0)$ 
for ingoing flows. 
For the polytropic equation of state with adiabatic index $\Gamma$, the 
relativistic specific enthalpy $\mu$ is written as (see Camenzind 1987),
\begin{equation}
   \mu  = m_{\rm p} \left[ 1 + h_{\rm inj}
         \left(\frac{u_p^{\rm inj} B_p}{u_p B_p^{\rm inj}}\right)
         ^{\Gamma-1}\right] \ ,                           \label{eq:mu}
\end{equation}
where 
\begin{equation}
    h_{\rm inj} \equiv \frac{\Gamma}{\Gamma-1}
                       \frac{P_{\rm inj}}{n_{\rm inj} m_{\rm p}}  \ , 
\end{equation}
and $m_{\rm p}$ is the rest mass of the particle. The quantities labeled 
by ``inj'' are specified at a point injecting plasma as a plasma source.  
The boundary conditions should be given by a plasma source model  
(e.g., the accretion disk/corona model, pair-creation model, and so on).  
Thus, we require the specification of the additional fifth constant of 
motion, $h_{\rm inj}$; in a cold limit, we obtain $h_{\rm inj}\to 0$ 
and $\mu\to m_{\rm p}\, (\equiv \mu_{\rm c})$.

By using the conserved quantities, the equation of motion projected 
onto the direction of a poloidal magnetic field, which is called the 
poloidal equation (and is often referred to as the relativistic 
Bernoulli equation), can be expressed by (e.g., Paper~I)
\begin{equation}
  (1+u_p^2) = (E/\mu)^2 \left[(\alpha -2M^2)f^2 - k \right] \ ,    
                                                       \label{eq:pol-eq}
\end{equation}
where
\begin{eqnarray}
  \alpha  &\equiv&  g_{tt}+2g_{t\phi}\Omega_F+g_{\phi\phi}\Omega_F^2  \ , \\
   k      &\equiv& (g_{\phi\phi}+2g_{t\phi}{\tilde L}+g_{tt}{\tilde L}^2)
                   /\rho_w^2 \ ,                           \label{eq:kk}  \\
   f      &\equiv&  -\ \frac{(g_{t\phi} + g_{\phi\phi}\Omega_F)
                  +(g_{tt} +g_{t\phi}\Omega_F){\tilde L}}
                             {\rho_w(M^2-\alpha )} \ , 
\end{eqnarray}
and $\tilde L\equiv L/E$. 
The relativistic Alfv\'en Mach-number $M$ is defined by 
\begin{equation}
   M^2 \equiv \frac{4\pi\mu n u_p^2}{B_p^2}=\frac{4\pi\mu\eta u_p}{B_p} \ . 
   \label{eq:Mach}
\end{equation}
Note that $\alpha^{-1/2}$ is the ``gravitational Lorentz factor'' of 
the plasma rotating with the angular velocity $\Omega_F$ in the Kerr 
geometry, whose definition includes both the effects of the
gravitational red-shift and the relativistic bulk motion in the 
toroidal direction.   
The locations of the Alfv\'en points $(r_{\rm A}, \theta_{\rm A})$ 
along a magnetic field line, where $\theta=\theta(r; \Psi)$, are 
defined by $M^2=\alpha$.

The relativistic force-balance equation, which is the equation of 
motion projected perpendicular to the magnetic surfaces, was derived 
by \citet{Nitta-TT91}; see also \citet{beskin97}. We should solve both 
the poloidal equation and the force-balance equation, but it is too 
difficult to solve these equations self-consistently. So, we only 
discuss the poloidal equation on a given magnetic field line. 
When the poloidal field geometry $\Psi=\Psi(r,\theta)$ is known and 
the various conserved quantities are specified at the injection point,  
the Mach-number (\ref{eq:Mach}) used together with equation(\ref{eq:mu}) 
in the poloidal equation (\ref{eq:pol-eq}) determines a complicated 
equation for $u_p$ as a function of $r$. It seems that, to obtain 
the poloidal  velocity, we must solve a polynomial of degree 
$4N+2$ in $z=u_p^{1/N}$ for the polytropic index $N=1/(\Gamma-1)$ 
(Camenzind 1987). 
To study the behavior of accretion and wind/jet solutions, however, 
we can reduce the poloidal equation to 
\begin{equation}
    E^2=\frac{\mu^2(1+u_p^2)(\alpha -M^2)^2}{{\cal R}},   \label{eq:fobid} 
\end{equation}
where ${\cal R} \equiv {\tilde e}^2\alpha -2{\tilde e}^2M^2-kM^4$ and 
${\tilde e} \equiv 1-\Omega_F\tilde L$, and we can plot a contour map of 
$(E/\mu_{\rm c})^2$ easily on the $r$-$u_p$ plane under the given flow 
parameters $\Omega_F$, $\tilde L$, $\eta$ and $h_{\rm inj}$. The regions 
with ${\cal R}<0$, where $E^2<0$ and no physical flow solution exists, 
give the forbidden regions of the flow solution on the $r$-$u_p$ plane; 
the shape of the forbidden regions are classified by $\Omega_F$ and 
$\tilde L$ (see Paper~I for the details of the forbidden regions).  
Note that, near the inner and outer light surfaces ($r=r_{\rm L}^{in}$ 
and $r=r_{\rm L}^{out}$) defined by $\alpha=0$, the toroidal velocity of 
the plasma approaches the light speed. However, if 
the ideal MHD condition breaks there because of the inertia effects, 
the plasma would enter the forbidden region $R<0$ by crossing the 
magnetic field lines; the non-ideal MHD flows are no longer forbidden 
in the $R<0$ regions. The breakdown of ideal MHD for accretion near the 
inner light surface is discussed in the last section.

At the Alfv\'en point ({\sf A}), where $r=r_{\rm A}$ and $M^2=\alpha$, 
it seems that the function $f$ diverges. 
However, to obtain a physical accretion solution passing through 
this point smoothly, we must require the following condition 
\begin{equation}
   {\tilde L}
    = \frac{(-g_{\phi\phi})_{\rm A}
      (\Omega_F-\omega_{\rm A})}{(g_{tt}+g_{t\phi}\Omega_F)_{\rm A}}\ ,
                    \label{eq:AP}
\end{equation}
where $\omega\equiv -g_{t\phi}/g_{\phi\phi}$ is the angular velocity 
of the zero angular momentum observer with respect to distant observers 
and the subscript ``A'' means the quantities at the Alfv\'en point. Thus, 
the ratio of the total angular momentum to the total energy of the flow 
is determined by the location of the Alfv\'en point and $\Omega_F$. 
When $0<\Omega_F<\omega_{\rm H}$ (classified as ``type II'' in Paper~I), 
where $\omega_{\rm H}$ is the angular velocity of the black hole, 
{\it one} Alfv\'en point {\sf A}$^{in}$ or {\sf A}$^{out}$ appears in 
the $r$-$u_p$ plane. 
When $\omega_{\rm H}<\Omega_F<\Omega_{\rm max}$ (``type I'') or 
$\Omega_{\rm min}< \Omega_F< 0$ (``type III''), {\it two} Alfv\'en 
points {\sf A}$^{in}$ and {\sf A}$^{in/out}$ appear, where 
$\Omega_{\rm min}$ and $\Omega_{\rm max}$ are the minimum and maximum 
angular frequency for the light surfaces to exist in the magnetosphere. 
If two Alfv\'en points appear in the $r$-$u_p$ plane, the physical flow 
solution passes through one of them. The Alfv\'{e}n point labeled by 
``{\it in}'' or ``{\it out}\/'' corresponds to inside or outside 
the separation point $r=r_{\rm sp}$ ({\sf SP}) which is defined by 
$\alpha'=0$. 
At $r=r_{\rm sp}$, the gravitational force and the centrifugal force 
on plasma are balanced if the poloidal velocity of the plasma is zero, 
hence the term separation point. 
Although the Alfv\'en point is classified into three types, one can 
interpret that the MHD flow of type~I, II or~III accretes onto a 
slow-rotating, rapid-rotating or counter-rotating black hole, 
respectively, when the value of $\Omega_F$ is specified.

From equations (\ref{eq:pol-eq}) and (\ref{eq:AP}), we can express 
the total energy $E$ and the total angular momentum $L$ as functions 
of the Alfv\'en radius and the injection point as follows:
\begin{eqnarray}
  E &=& \frac{(g_{tt}+g_{t\phi}\Omega_F)_{\rm A}} 
             {\alpha_{\rm A}} {\cal E}                \ ,  \label{eq:E}\\
  L &=& \frac{(-g_{\phi\phi})_{\rm A}(\Omega_F-\omega_{\rm A})}
             {\alpha_{\rm A}} {\cal E}                \ ,  \label{eq:L}  
\end{eqnarray}
where ${\cal E} \equiv E-\Omega_F L > 0 $ can be specified at the plasma 
injection point and the Alfv\'en point. 
The condition of a negative energy MHD accretion flow, 
$ (g_{tt}+g_{t\phi}\Omega_F)_{\rm A}<0 $, is unchanged from the cold 
limit case (see Paper~I). Thermal effects on the hot plasma flow are 
only included in ${\cal E}$, and modify the amplitude of the ingoing 
energy and angular momentum.

\section{Critical Conditions at Magnetosonic Points } 

For a physical solution for accretion onto a black hole, we also require 
that the critical conditions at both fast and slow magnetosonic points 
must be satisfied. In this section, we discuss restrictions on the
remaining field-aligned parameters; we see that the locations of the 
fast and slow magnetosonic points give the total energy and the particle 
number flux along the magnetic field lines.

The differential form of the poloidal equation (\ref{eq:pol-eq}) is 
written by (see Paper~I)
\begin{equation}
   (\ln u_p)' = \frac{{\cal N}}{{\cal D}}\ ,            \label{eq:ND}
\end{equation}
where
\begin{eqnarray}
   {\cal N} &=& \left(\frac{E}{\mu}\right)^2 \left\{
         \left[{\cal R}(M^2-\alpha )C_{\rm sw}^2 + M^4{\cal A}^2 \right] 
         (\ln B_p)' 
         +\frac{1}{2}(1+C_{\rm sw}^2) \left[M^4(M^2-\alpha )k' 
         -{\cal Q}\alpha ' \right] \right\}     \ ,     \label{eq:numer}\\
   {\cal D} &=& (M^2-\alpha )^2\left[(C_{\rm sw}^2-u_p^2)(M^2-\alpha ) 
         +(1+u_p^2) M^4 {\cal A}^2{\cal R}^{-1}
         \right] \                                       \label{eq:domom}
\end{eqnarray}
with 
${\cal A}^2 \equiv {\tilde e}^2+\alpha k = f^2(M^2-\alpha)^2$ and 
${\cal Q}   \equiv \alpha {\tilde e}^2 - 3{\tilde e}^2 M^2 - 2k M^4 $. 
The prime $(\cdots )'$ denotes 
$[(\partial_\theta\Psi)\partial_r -(\partial_r\Psi)\partial_\theta]/
(\sqrt{-g}B_p)$ which is a derivative along a stream line. 
The relativistic sound velocity $a_{\rm sw}$ is given by 
\begin{equation}
   a_{\rm sw}^2 \equiv \left(\frac{\partial\ln \mu}
                                  {\partial\ln n} \right)_{\rm ad} 
          = (\Gamma-1)\frac{\mu-m_{\rm p}}{\mu} \ , 
\end{equation}
and the sound four-velocity is given by 
$C_{\rm sw}^2 = a_{\rm sw}^2/(1-a_{\rm sw}^2)$.

The denominator (\ref{eq:domom}) can be reduced to the form 
\begin{equation}
    {\cal D} \propto \left(u_p^2-u_{\rm AW}^2\right)^2
              \left(u_p^2-u_{\rm FM}^2\right)
              \left(u_p^2-u_{\rm SM}^2\right)\ , 
\end{equation}
where the relativistic Alfv\'en wave speed $u_{\rm AW}$, the fast 
magnetosonic wave speed $u_{\rm FM}$ and the slow magnetosonic wave 
speed $u_{\rm SM}$ are defined by 
\begin{eqnarray}
   u_{\rm AW}^2(r;\Psi) &\equiv & \frac{B_p^2}{4\pi\mu n}\alpha  \ , 
                                                            \label{eq:aw}\\
   u_{\rm FM}^2(r;\Psi) &\equiv & \frac{1}{2} \left({\cal Z} 
         + \sqrt{{\cal Z}^2 -4 C_{\rm sw}^2 u_{\rm AW}^2 }\right)\ , 
                                                            \label{eq:fw}\\
   u_{\rm SM}^2(r;\Psi) &\equiv & \frac{1}{2} \left({\cal Z} 
         - \sqrt{{\cal Z}^2 -4 C_{\rm sw}^2 u_{\rm AW}^2 }\right)\ , 
                                                            \label{eq:sw} 
\end{eqnarray}
with 
\begin{eqnarray}
    {\cal Z} &\equiv & 
       u_{\rm AW}^2 + \frac{B_\phi^2}{4\pi\mu n \rho_w^2}+C_{\rm sw}^2\ , \\
    B_\phi &=& - 4\pi\eta E \rho_w f \ .  
\end{eqnarray}
When $u_p^2=u_{\rm AW}^2$, $u_p^2=u_{\rm FM}^2$ or $u_p^2=u_{\rm SM}^2$, 
the denominator becomes zero. Therefore, at these singular points, we must 
require ${\cal N}=0$ to obtain physical accretion solutions which pass 
through these points smoothly. The location of 
$u_p^2=u_{\rm A}^2$ [$\equiv u_{\rm AW}^2(r_{\rm A};\Psi)$] is the Alfv\'en 
point discussed in the previous section. Similarly, the locations of 
$u_p^2=u_{\rm F}^2$ [$\equiv u_{\rm FM}^2(r_{\rm F};\Psi)$] and 
$u_p^2=u_{\rm S}^2$ [$\equiv u_{\rm SM}^2(r_{\rm S};\Psi)$] correspond 
to the fast magnetosonic point $r=r_{\rm F}$ and the slow magnetosonic 
point $r=r_{\rm S}$, respectively. 
We should mention that, to calculate the Alfv\'en velocity, 
we need to solve a polynomial of high degree, while in the 
cold limit it is simply obtained as 
$u_{\rm A}^{\rm cold}=(\alpha  B_p)_{\rm A}/(4\pi\mu_{\rm c} \eta)$; 
the Alfv\'en velocity of a hot MHD flow is always smaller than 
$u_{\rm A}^{\rm cold}$.

Figure~\ref{fig:dd} shows schematic pictures of the ${\cal D}=0$ curves 
in the $r$-$u^r$ plane. [The definition of the poloidal velocity $u_p$ 
includes the gravitational-redshift factor, so that $u_p$ diverges at 
the event horizon for a physical accretion solution. Hereafter we use 
the $r$-$u^r$ plane when we discuss the behavior of accretion solutions; 
under a given magnetic field with $\Psi(r, \theta)=$ constant, we can also 
calculate $\theta=\theta(r;\Psi)$ and $u^\theta=u^\theta(r;\Psi)$. ]\ 
Corresponding to the two modes of magnetosonic wave speeds and the 
Alfv\'en wave speed, we see three branches of ${\cal D}=0$ curves, 
which correspond to $u_p^2=u_{\rm FM}^2$, $u_p^2=u_{\rm SM}^2$ and 
$u_p^2=u_{\rm AW}^2$ curves. The $u_p^2=u_{\rm AW}^2$ curve is always 
located inside the forbidden region (the shaded regions) except for 
the Alfv\'{e}n point, so these forbidden regions separate the 
$r$-$u^r$ plane into one or two super-Alfv\'{e}nic region(s) and 
one sub-Alfv\'{e}nic region. 
If there is no area of $k(r;\tilde{L})>0$, we see one 
super-Alfv\'enic and one sub-Alfv\'enic regions with forbidden regions 
classified as  ``type~A'' in Paper~I. The total angular momentum has 
a value of $\tilde{L}_{-} < \tilde{L} < \tilde{L}_{+}$, where 
$\tilde{L}=\tilde{L}_{+} (>0)$ and $\tilde{L}=\tilde{L}_{-} (<0)$ 
are the minimum and maximum values of the function 
$\tilde{L}=\tilde{L}(r_{\rm A};\Omega_F)$. 
On the other hand, if an area with $k>0$ exists between two light 
surfaces, we see two super-Alfv\'enic regions separated by ``type~B'' 
forbidden regions, and obtain larger total angular momentum MHD flows 
($\tilde{L} > \tilde{L}_{+}$ or $|\tilde{L}| > |\tilde{L}_{-}|$). 
Thus, we can classify the forbidden regions by $\Omega_F$ and $\tilde L$ 
as type~I, II or~III and type~A or~B, independently; hereafter, we will 
denote the type of forbidden regions as, for example, type~IA.

\placefigure{fig:dd}

The $u_p^2=u_{\rm FM}^2$ and $u_p^2=u_{\rm SM}^2$ curves are located 
in the super-Alfv\'{e}nic region and the sub-Alfv\'{e}nic region, 
respectively. Figure~\ref{fig:dd}a shows a case for strong magnetic 
fields satisfying $(C_{\rm sw}^{2})_{\rm A} < u_{\rm A}^2$, and 
Figure~\ref{fig:dd}b shows a case for weak magnetic fields satisfying 
$(C_{\rm sw}^{2})_{\rm A} > u_{\rm A}^2$. In Figure~\ref{fig:dd}a the 
$u_p^2=u_{\rm FM}^2$ curve connects to the Alfv\'{e}n point marked by 
``{\sf A}'' ($r=r_{\rm A}, u_p^2=u_{\rm A}^2$), while in 
Figure~\ref{fig:dd}b the $u_p^2=u_{\rm SM}^2$ curve connects to the 
Alfv\'{e}n point. At the Alfv\'{e}n {\it radius}\/ $r=r_{\rm A}$ the 
value of the function $f=f(u_p; r_{\rm A})$ is zero except at the 
Alfv\'en point {\sf A}. Then, $u_p^2=C_{\rm sw}^2$ is one of the 
solutions of ${\cal D}(u_p; r_{\rm A})=0$ (marked by ``{\sf C}'' in 
Fig.~\ref{fig:dd}). A similar situation at the (outer) Alfv\'en point 
can be seen in the Newtonian case (see \citet{Heyvaerts-Norman89}).

\placefigure{fig:nn}

We also plot a typical curve with ${\cal N}=0$ in 
Figure~\ref{fig:nn}. Crossing of the ${\cal D}=0$ curves and 
${\cal N}=0$ curves in the super- or sub-Alfv\'enic region means 
the fast or slow magnetosonic point, respectively. In the cold limit, 
between the Alfv\'en point and the event horizon in the super-Alfv\'enic 
region, crossing of the ${\cal N}=0$ curve and ${\cal D}=0$ curve 
always exists regardless of the $\eta$ value (Paper~I). However, 
in the case of $(C_{\rm sw}^{2})_{\rm A}>u_{\rm A}^2$, we cannot find 
any reason for crossing of these lines. In fact, we find a restriction 
on the hot trans-fast MHD accretion by the thermal effects. In this case, 
the condition ${\cal N}={\cal D}=0$ is not achieved between the inner 
Alfv\'en radius and the event horizon; that is, no physical trans-fast 
MHD accretion solution exists. When the thermal effects dominate over 
the magnetic effects, we can find that the crossing of these lines is 
only available for smaller $|\eta|$, while for larger $|\eta|$ it 
becomes impossible to generate a physical trans-fast MHD accretion 
solution (see below). 
In the cold limit, because the slow magnetosonic wave speed is zero, 
the X-type slow magnetosonic point is located just on the $r$-axis 
($u^r=0$ line), which is just the separation point. 
When thermal effects are effective, we can see X-type slow 
magnetosonic points with a sub-slow magnetosonic region in the 
sub-Alfv\'enic region of the $r$-$u^r$ plane.

Now, we will discuss the condition for crossings of the 
${\cal N}(r,u_p)=0$ and ${\cal D}(r,u_p)=0$ curves. We use the indices 
``F'' and ``S'' to denote the quantities evaluated at the fast and slow 
magnetosonic points, respectively, and use the index ``cr'' to unite  
the quantities at these magnetosonic points.  In the following equations, 
to discuss MHD flows passing through the fast or slow magnetosonic point, 
we can replace the subscript ``cr'' by ``F'' or ``S''. 
From the condition ${\cal D}=0$ at the fast and slow magnetosonic points, 
the poloidal velocity at these critical points 
$u_{\rm cr}^2\equiv u_p^2(r_{\rm cr}; \Psi)$ is written as 
\begin{equation}
    u_{\rm cr}^2 
       = \left[\frac{{\cal R}C_{\rm sw}^2(M^2-\alpha )+M^4{\cal A}^2}
                   {{\cal R}(M^2-\alpha )-M^4{\cal A}^2} 
        \right]_{\rm cr} \ , 
\end{equation}
and by use of the definition of Mach-number (\ref{eq:Mach}), 
the particle flux through a flux tube $\eta$ is determined by 
\begin{equation}
  \eta = \left(\frac{B_p}{4\pi\mu}\right)_{\rm cr}
         \frac{M^2_{\rm cr}}{u_{\rm cr}} \ , 
         \label{eq:eta}
\end{equation}
where the critical Mach-number $M_{\rm cr}^2\equiv M^2(r_{\rm cr}; \Psi)$ 
is obtained as a solution of ${\cal N}=0$, which is a cubic equation in 
$M^2$.  Thus, we can express $\eta$ as a function of $r_{\rm cr}$ with 
given parameters $\Omega_F$, $\tilde L$, $(a_{\rm sw}^2)_{\rm cr}$ and 
$\Psi$. The total energy of the trans-fast (or trans-slow) MHD flow is 
also evaluated at the fast (or slow) magnetosonic point $r_{\rm F}$ 
(or $r_{\rm S}$) by using the poloidal equation (\ref{eq:fobid}).  
Here, we would like to emphasize that $(a_{\rm sw}^2)_{\rm cr}$   
is introduced as a parameter for the thermal effects on the trans-fast 
MHD flows instead of $(a_{\rm sw}^2)_{\rm inj}$ (or $h_{\rm inj}$). 
The acceptable ranges for $(a_{\rm sw}^2)_{\rm S}$ and 
$(a_{\rm sw}^2)_{\rm F}$ are restricted by the critical condition 
(\ref{eq:eta}), which is to be realized as trans-slow and trans-fast 
MHD accretion, respectively. Thus, all boundary conditions can be 
replaced by the Alfv\'en and magnetosonic conditions. 
The behavior of 
$\eta = \eta[r_{\rm cr}; \Omega_F, \tilde L, (a_{\rm sw}^2)_{\rm cr}; 
a, \Psi]$ will be discussed in the next section.

\section{Thermal Effects on Trans-Fast MHD Flows} 

Let us discuss a trans-fast MHD flow in a black hole magnetosphere. 
We consider magnetic flux-tubes given by $\partial_r\Psi=0$ and  
$\partial_\theta\Psi = C\,r^{-\delta}\,\sin\theta$; that is, 
the poloidal magnetic field is denoted as 
$B_p(r; \Psi)=(C/\sqrt{\Delta\Sigma}) \,r^{-\delta}$, 
where $\delta=\delta(\Psi)$ and $C=C(\Psi)$; 
$\delta$ is related to the divergence of a 
magnetic flux tube. For example, $\delta=0$ with $C(\Psi)=$ constant 
means the split monopole magnetic field \citep{BZ77}. 
Hereafter, we consider a situation where the plasma streams close to 
the equatorial plane. We expect that the qualitative picture is not 
drastically changed when we leave the equatorial plane and when we 
consider more complicated field geometries.

\subsection{Restrictions on Trans-Fast MHD Flow Solutions} 

We introduce a new parameter $x_{\rm A}$ to specify the Alfv\'en radius by 
$x_{\rm A}\equiv (r_{\rm sp}-r_{\rm A})/(r_{\rm sp}-r_{\rm L}^{in})$, 
where $0<x_{\rm A}<1$. Though, for a given $x_{\rm A}$, we obtain a value 
of $\tilde L$, we may find another value for $x_{\rm A}$ giving the same 
$\tilde L$ value in the cases with types~I and~III; that is, we may see 
two Alfv\'en points inside the separation point. 
Further, we introduce $\hat\eta \equiv \mu_{\rm c}|\eta/C|$ and 
$\zeta_{\rm cr}\equiv (a_{\rm sw}^2)_{\rm cr}$.

\placefigure{fig:eta}

Figures~\ref{fig:eta}a,~\ref{fig:eta-OM}--\ref{fig:eta-typeII} 
show relations between $\hat\eta$ and $r_{\rm cr}$ for various 
$\zeta_{\rm cr}$ values under given parameters $\Omega_F$, $\tilde L$, 
$\delta$ and the spin parameter $a$; The locations of 
$r_{\rm cr}=r_{\rm H}$, $r_{\rm cr}=r_{\rm L}^{in/out}$, 
$r_{\rm cr}=r_{\rm A}^{in/out}$ and $r_{\rm cr}=r_{\rm sp}$ are marked 
by {\sf H}, {\sf L}, {\sf A} and {\sf SP}, respectively. 
Here, we will discuss the magnetosonic points located inside the outer 
light surface, because for accretion the fast magnetosonic point should 
be located inside the outer Alfv\'en point and the slow magnetosonic 
point should be located between the inner Alfv\'en point and the outer 
light surface.

\subsubsection{General Properties}

In the cold limit, 
there are three branches of $\hat\eta=\hat\eta(r_{\rm F})$ 
(solid curves with $\zeta_{\rm cr}=0.0$), while there is no 
$\hat\eta=\hat\eta(r_{\rm S})$ curve. For hot MHD flows with 
$\zeta_{\rm cr}=0.1, 0.2$ and $0.3$, however, both 
$\hat\eta=\hat\eta(r_{\rm F})$ and $\hat\eta=\hat\eta(r_{\rm S})$ 
curves (dashed curves) exist. When we try to plot a contour map of 
$(E/\mu_{\rm c})^2$ on the $r$-$u^r$ plane as an accretion solution 
(see, e.g., Fig.~\ref{fig:MHDaccA}a), from the $\hat\eta$ 
vs. $r_{\rm cr}$ diagram we can find the acceptable locations of 
the fast/slow magnetosonic points, which are shown as the crossings 
of a $\hat\eta=$ constant line and 
$\hat\eta=\hat\eta(r_{\rm cr}; \zeta_{\rm cr})$ curves.  
Hereafter, the magnetosonic points located inside the inner Alfv\'en 
point are labeled as ``in'', the middle magnetosonic points located 
between two Alfv\'en points are labeled as ``mid'' and the magnetosonic 
points located outside the outer Alfv\'en point are labeled as ``out''.
In Figures~\ref{fig:eta}a,~\ref{fig:eta-OM}--\ref{fig:eta-typeII}, 
the location of $\hat\eta(r_{\rm cr}; \zeta_{\rm cr})=0$ exists between 
$r_{\rm F}=r_{\rm H}$ and $r_{\rm F}=r_{\rm A}^{in}$; The location 
does not depend on the $\zeta_{\rm F}$ values. For hot MHD flows 
we can see the cases where $\hat\eta=\hat\eta(r_{\rm F})$ 
curves and $\hat\eta=\hat\eta(r_{\rm S})$ curves are connected at 
$r_{\rm cr}=r_{\rm A}$ (the locations marked by ``$\bullet$'').  
For the fast magnetosonic point located just inside (or outside) the 
Alfv\'en point, from ${\cal N}=0$ we have 
\begin{equation}
    M_{\rm F}^2 
    = \alpha_{\rm A}+{\cal Y}_{\rm A}|{\cal A}|^{1/2}+O({\cal A})
\end{equation}
where
\begin{equation}
    {\cal Y} \equiv \frac{{\cal A}}{|{\cal A}|}
             \frac{\{ 6{\tilde e}^2[a_{\rm sw}^2(\ln B_p)']^2
                 -2{\tilde e}^2 a_{\rm sw}^2(\ln B_p)'{\cal A}'
                 +k\alpha '{\cal A}' \}^{1/2}} 
                 {-k a_{\rm sw}^2(\ln B_p)' +(k/2)'} \ . 
\end{equation}
In the $r_{\rm F}\to r_{\rm A}$ (${\cal A}\to 0$) limit, 
from equation (\ref{eq:eta}), the value of $\eta$ is given by 
\begin{eqnarray}
  \hat\eta_\ast 
  &\equiv& \lim_{r_{\rm F}\to r_{\rm A}} \hat\eta            \nonumber  \\
  &=& \left[ \frac{(B_p/C)\alpha }{4\pi(\mu/\mu_{\rm c})} \right]_{\rm A}
      \left[ \frac{ k \{ 6{\tilde e}^2 [a_{\rm sw}^2(\ln B_p)']^2 
        -2{\tilde e}^2 a_{\rm sw}^2 (\ln B_p)'{\cal A}' 
              +k \alpha' {\cal A}' \} }
        {a_{\rm sw}^2 \{[2ka_{\rm sw}^2(\ln B_p)'-k' ] \alpha k{\cal A}' 
        +3{\tilde e}^2 a_{\rm sw}^2 (\ln B_p)'k'\} } - 1  
       \right]^{1/2}_{\rm A} \ . 
\end{eqnarray}
Then, $\hat\eta$ for hot MHD accretion passing through the 
inner-fast magnetosonic point has an upper-limit.
Thus, the thermal effect restricts the acceptable $\hat\eta$ values, 
while in the cold limit the range is given by $0 < \hat\eta < \infty$. 
Figures~\ref{fig:eta}a,~\ref{fig:eta-OM}--\ref{fig:eta-typeII} 
also show that the $\hat\eta=\hat\eta(r_{\rm F}^{\rm in})$ and 
$\hat\eta=\hat\eta(r_{\rm F}^{\rm out})$ curves shrink down vertically 
with increasing $\zeta_{\rm F}$. 
When $\zeta_{\rm F}\geq 0.3$, the magnetic effect on the plasma can still 
remain efficient for the flows passing through $r_{\rm F}^{\rm in}$ 
or $r_{\rm F}^{\rm out}$ which gives a trans-fast MHD flow with  
$\hat\eta \ll 1$. 
On the other hand, the $\hat\eta=\hat\eta(r_{\rm S}^{\rm in})$ and 
$\hat\eta=\hat\eta(r_{\rm S}^{\rm out})$ curves become almost 
vertical when $\hat\eta$ is at least several times as large as 
$\hat\eta_\ast$; that is, the location of the slow magnetosonic point 
is almost independent of $\hat\eta$, and the curves shift toward the 
inner and outer light surfaces with increasing $\zeta_{\rm S}$, 
respectively. 
We should note that an ideal MHD accretion flow after passing through the 
inner-slow magnetosonic point is impossible, because no Alfv\'en point 
is located inside this slow magnetosonic point; an outflow may be possible 
after passing through this slow magnetosonic point, the inner Alfv\'en 
point and the middle fast magnetosonic point, in this order.

In contrast to the inner-magnetosonic points, a trans-fast MHD flow 
passing through the middle-fast magnetosonic point 
is always available for any $\hat\eta$ values. 
We see that in Figure~\ref{fig:eta}a the location of the middle-fast 
magnetosonic point is insensitive to both $\hat\eta$ and 
$\zeta_{\rm F}$; and the location of the middle-slow magnetosonic 
point is insensitive to only $\hat\eta$, while it shifts inward with 
increasing $\zeta_{\rm S}$. 
We can also see branches for $\hat\eta=\hat\eta(r_{\rm F}^{\rm out})$ 
for the fast magnetosonic point located outside the outer Alfv\'en point. 
An outgoing flow from the black hole magnetosphere should pass through 
the outer-fast or middle-fast magnetosonic point after passing through 
the outer or inner Alfv\'en point. In this paper, however, we will focus 
our attention to trans-fast MHD accretion onto a black hole, and omit 
discussions of the outgoing flows from the magnetosphere.

Figure~\ref{fig:eta}b shows the total energy $E/\mu_{\rm c}$ of the 
trans-magnetosonic flows as a function of $r_{\rm cr}$. For a fast 
magnetosonic point giving a smaller $\hat\eta$ value, the energy 
becomes larger. This means that the larger energy is shared by smaller 
numbers of particles. 
Corresponding to the ``maximum'' $\hat\eta$ value for 
$\hat\eta=\hat\eta(r_{\rm F}^{\rm in})$, a ``minimum'' value of 
$E/\mu_{\rm c}$ exists for the $E=E(r_{\rm F}^{\rm in/out})$ curve. 
For a hotter flow which includes more thermal energy, this minimum 
value becomes larger than the cooler one. A trans-slow MHD flow passing 
through the inner-slow or outer-slow magnetosonic point has also a lower 
limit for the total energy. Furthermore, for the middle-slow magnetosonic 
point, the acceptable value of $E/\mu_{\rm c}$ is restricted within a 
very narrow range, while it seems that there is no restriction on 
$\hat\eta$.

\subsubsection{Rotational Effects of the Magnetic Field Line}

From Figures~\ref{fig:eta-OM} and~\ref{fig:eta}a, we can see the 
$\Omega_F$ dependence of the locations of magnetosonic points. 
Comparing Figure~\ref{fig:eta-OM}a ($\Omega_F=0.9\Omega_{\rm max}$) 
with Figure~\ref{fig:eta}a ($\Omega_F=0.8\Omega_{\rm max}$), the 
location of the inner light surface moves outward and the locations 
of the separation point and the outer light surface move inward with 
increasing $\Omega_F$.  
In both cases, the separation point is located between two Alfv\'en 
points. For cooler accretion of $\zeta_{\rm F}\leq 0.1$, three 
fast magnetosonic points may be possible between the inner Alfv\'en 
point and the event horizon (see Fig.~\ref{fig:eta-OM}a).  
The middle one is an O-type point (unphysical), while the others are 
X-type critical points (physical). 
For hotter accretion ($\zeta_{\rm F} \geq 0.1$ at least), however, 
the X-type fast magnetosonic point located next to the Alfv\'en point 
disappears, and it turns to the slow magnetosonic point.

\placefigure{fig:eta-OM}

Next, comparing Figure~\ref{fig:eta-OM}b ($\Omega_F=0.5\Omega_{\rm max}$) 
with Figure~\ref{fig:eta}a, the location of the inner light surface 
moves inward and the locations of the separation point and the outer 
light surface move outward with decreasing $\Omega_F$. Two Alfv\'en 
points are located between the inner light surface and the separation 
point; and then, both the inner and outer Alfv\'en points, which are 
labeled by ``{\it in}'' (see \S~2), 
can be related to accretion started near the separation point. 
Concerning the $\hat\eta=\hat\eta(r_{\rm F}^{\rm mid})$ branches, 
which are located between two Alfv\'en points, a branch of smaller 
$\zeta_{\rm F}$ (e.g., $\zeta_{\rm F}=0.1$) has a maximum, while 
a branch of larger $\zeta_{\rm F}$ (e.g., $\zeta_{\rm F}=0.2, 0.3$) 
has no upper limit. We see that for smaller $\zeta_{\rm F}$ the 
branch for the middle-fast magnetosonic point connects to the outer 
Alfv\'en point (see the $\zeta_{\rm F}=0.1$ curve) and for larger 
$\zeta_{\rm F}$ the branch for the outer-fast magnetosonic point 
connects to the outer Alfv\'en point (see $\zeta_{\rm F}=0.2$).

Figure~\ref{fig:eta-alf} shows $\hat\eta$ as a function of $r_{\rm cr}$ 
with $x_{\rm A}=0.5$. 
The outer Alfv\'en point is located inside the separation point. 
There is an upper limit to the $\hat\eta=\hat\eta(r_{\rm F}^{\rm mid})$ 
curve for each of the $\zeta_{\rm F}=0.1$ and $0.2$ cases, while 
for $\zeta_{\rm F}=0.3$ the middle-fast magnetosonic point always 
exists for any $\hat\eta$ values (no upper limit).

\placefigure{fig:eta-alf}

\subsubsection{Black Hole's Spin Effects}

Figures~\ref{fig:eta-spin}a and~\ref{fig:eta-spin}b show the $\hat\eta$ 
vs. $r_{\rm cr}$ relation with $a=0.5m$ (a corotating black hole with 
the magnetosphere) and $a=-0.5m$ (a counter-rotating black hole), 
respectively. They can be also compared with Figure~\ref{fig:eta}a, 
which is the case with $a=0$. In the case with $a=0.5m$, two Alfv\'en 
points are located inside the separation point; and the outer Alfv\'en 
point is located very close to the separation point. The properties of 
$\hat\eta=\hat\eta(r_{\rm cr})$ are similar to the case of 
Figure~\ref{fig:eta-OM}b. In Figure~\ref{fig:eta-spin}a, it seems that the 
branches with $\hat\eta=\hat\eta(r_{\rm F}^{\rm mid})$ have no maximum 
value, but for cooler flows ($\zeta_{\rm F}\ll 1$) there are 
branches with $\hat\eta=\hat\eta(r_{\rm F}^{\rm mid})$ which have a 
maximum value. 
In the case of $a=-0.5m$, the separation point is located between two 
Alfv\'en points, and the properties of $\hat\eta=\hat\eta(r_{\rm cr})$ 
are essentially similar to the cases for Figures~\ref{fig:eta}a  
and~\ref{fig:eta-OM}a. Comparing Figure~\ref{fig:eta-spin}b with 
Figure~\ref{fig:eta}, the counter-rotating effect of the black hole 
generates three (or two) inner-fast magnetosonic points, while the 
effect of corotation is to suppress such multi-inner-fast magnetosonic 
points generation (see, e.g., $\hat\eta=\hat\eta(r_{\rm F}^{\rm in})$ 
curves with $\zeta_{\rm F}=0.1$ in these figures).

\placefigure{fig:eta-spin}

\placefigure{fig:eta-typeII}

Figure~\ref{fig:eta-typeII} shows $\hat\eta$ as a function of $r_{\rm cr}$ 
with $a=0.8m$, $x_{\rm A}=0.8$ and $\Omega_F=0.5\Omega_{\rm max}$.  
This is a case of only one Alfv\'en point and one super-Alfv\'enic  
region in the $r$-$u^r$ plane (type IIA forbidden region). In the 
previous examples of type IA forbidden region 
(i.e., Figs.~\ref{fig:eta}--\ref{fig:eta-typeII}), we have seen inner 
and outer Alfv\'en points and two regimes of trans-fast MHD accretion 
solutions. In this case, however, there is one type of solution. The 
Alfv\'en point is located inside the separation point, and each branch 
of $\hat\eta=\hat\eta(r_{\rm F}^{\rm in})$ has an upper limit for 
$\hat\eta$, except for the cold limit.  
It seems that for the cooler flows the slow magnetosonic point is 
located near the separation point, but for the hotter flows the location 
shifts outward. However, if the hotter flow has a small $\hat\eta$, the 
slow magnetosonic point could be located close to the separation point.

\subsubsection{Non-conical Effects of the Magnetic Field Geometry}

The effects of magnetic field geometry on the critical condition 
(\ref{eq:eta}) are shown in Figure~\ref{fig:delta} for a hot MHD flow;  
The effects of non-conical geometry ($\delta\neq 0$) for cold  
MHD accretion have been discussed by \citet{Takahashi94}. 
The maximum $\hat\eta$ value increases with increasing $\delta$, 
for $\hat\eta=\hat\eta(r_{\rm F}^{\rm in})$ curves. 
This means that the field geometry converging along an ingoing stream 
line ($\delta>0$) rather than the radial field is available to make 
a higher accretion-rate than $\delta<0$ cases. 
The location of $\hat\eta(r_{\rm F})=0$ has a weak dependence on 
$\delta$. The location of the middle fast magnetosonic point also has 
a weak dependence on $\delta$, and the slow magnetosonic point remains 
at a fixed location except for smaller $\hat\eta$ cases.

\placefigure{fig:delta}

\subsection{Two Regimes of Accretion Solutions } 

Here, we will present accretion solutions. 
To plot an {\sf SAF}-solution, we need to determine the five 
field-aligned constant quantities: $\Omega_F$, $L$, $\eta$, $E$, 
$h_{\rm inj}$. Although these constants should be given as boundary 
conditions at the plasma source ({\sf inj}), mathematically we can 
choose the locations 
of $r_{\rm F}$, $r_{\rm L}$, $r_{\rm A}$ and the sound velocity 
$\zeta_{\rm F}$ as free parameters to fix the conserved quantities 
if one restricts their interest to ideal MHD solutions.  
First, for the plasma sources to exist in a black hole magnetosphere, 
we require that $\Omega_{\rm min}<\Omega_F<\Omega_{\rm max}$; then, 
two light surfaces $r=r_{\rm L}^{in}(\Omega_F; a,\Psi)$ and 
$r=r_{\rm L}^{out}(\Omega_F; a,\Psi)$ are determined. 
Second, from equation (\ref{eq:AP}), $\tilde L$ is determined by the
Alfv\'en radius $r_{\rm A}$, which is located between two light 
surfaces mentioned above. 
Third, as we have seen in the previous section, by specifying 
$r_{\rm F}$ and $\zeta_{\rm F}$, $\eta$ and $E$ are calculated from 
equations (\ref{eq:eta}) and (\ref{eq:fobid}). Similarly, when 
$r_{\rm S}$ and $\zeta_{\rm S}$ are set, $\eta$ and $E$ are also 
calculated. Of course, we should require that 
$\eta=\eta(r_{\rm F}, \zeta_{\rm F})=\eta(r_{\rm S}, \zeta_{\rm S})$ 
for the {\sf SAF}-solution. 
Finally, we can plot a contour map of $E/\mu_{\rm c}$ on the $r$-$u^r$ 
plane. The {\sf SAF}-solution is obtained as the curve with 
$E=E_{\rm F}=E_{\rm S}$, where 
$E_{\rm cr}\equiv E(r_{\rm cr}, \zeta_{\rm cr})$.

\placefigure{fig:MHDaccA}

\placefigure{fig:MHDaccB}

When the forbidden region is type IA (or IIIA), where both inner and 
outer Alfv\'{e}n points appear on the $r$-$u^r$ plane, there are two 
regimes of {\sf SAF}-MHD accretion solutions (see \cite{Takahashi00}): 
\begin{description}
 \item[(i)] 
  ${\sf inj} \to {\sf S}^{\rm mid} \to {\sf A}^{in} \to 
   {\sf F}^{\rm in} \to {\sf H}$ 
 \item[(ii)] 
  ${\sf inj} \to {\sf S}^{\rm out} \to {\sf A}^{out} \to 
   {\sf F}^{\rm mid} \to {\sf H}$ 
\end{description}
The accreting matter for case (i) would be injected near the separation 
point. In Figure~\ref{fig:eta}b, we see that the energy 
$E_{\rm S}/\mu_{\rm c}$ is restricted within a very narrow range.  
Though the sound velocity is not constant along the flow, which means 
$\zeta_{\rm S}\neq\zeta_{\rm F}$, the possible location of the 
inner-fast magnetosonic point which must give $E_{\rm F}=E_{\rm S}$ 
would be also restricted to a narrow range. 
Note that if the accreting plasma is intensely heated up, there 
may be no solution of case (i); for example, in Figure~\ref{fig:eta}b, 
the flow of $\zeta_{\rm S}=0.1$ and $\zeta_{\rm F}=0.2$ is forbidden, 
because such plasma heating causes a conflict situation of 
$E_{\rm F} > E_{\rm S}$.  
The accreting matter for case (ii) would be injected inward from 
an area between the outer Alfv\'en radius and the outer light surface. 
If the separation point is located outside the outer Alfv\'en point, 
it is possible that the case (ii) accreting matter is injected from 
near the separation point. 
Figures~\ref{fig:MHDaccA} and~\ref{fig:MHDaccB} show typical examples 
of case (i) and case (ii), which satisfy the requirement that 
$\hat\eta < \hat\eta_{\rm max}(r_{\rm F}^{\rm in})$. 
In Figures~\ref{fig:MHDaccA}a and~\ref{fig:MHDaccB}a, we also see 
accretion solutions which reach the event horizon with zero radial 
velocity, but these solutions are unphysical because the accreting 
plasma stops just on the event horizon and its density diverges.

Now, we introduce $X_{\rm em}(r) \equiv -\Omega_F B_\phi/(4\pi\eta |E|)$, 
which means the ratio of the electromagnetic energy to the total 
energy (absolute value); by normalizing with $|E|$, we will express 
$X_{\rm em}<-1$ for a negative energy ($E<0$) inflow.  
The fluid part of energy per total energy 
is denoted by $E/|E|-X_{\rm em} (\equiv X_{\rm fluid})$, where 
$(X_{\rm fluid})_{\rm H}$ can become negative even if 
$(X_{\rm fluid})_{\rm inj}$ is positive \citep{Hirotani-TNT92}; that is, 
the initial positive energy $(\mu u_t)_{\rm inj}$ is extracted from the 
plasma and is deposited in the magnetic field to be carried outwards. 
Figures~\ref{fig:MHDaccA}b and~\ref{fig:MHDaccB}b show the energy  
conversion between the fluid part and electromagnetic part in the 
Schwarzschild geometry, where $X_{\rm fluid}$ is always positive. 
Though the poloidal flow solution in the black hole magnetosphere 
contains two Alfv\'en radii $r=r_{\rm A}^{in}$ and 
$r=r_{\rm A}^{out}$, an accretion across both Alfv\'en radii is 
possible when the injection point is located between the outer 
Alfv\'en point and the outer light surface. 
One of them corresponds to the Alfv\'en point for the considered flow, 
where the requirement of $u_p^2=u_{\rm AW}^2$ is satisfied, and 
$X_{\rm em}$ does not change its sign; such a point is {\sf A}$^{in}$ 
for case (i) and {\sf A}$^{out}$ for case (ii). However, $X_{\rm em}$ 
changes its sign at the other Alfv\'en {\it radius}, which is not the 
Alfv\'en point for the considering {\sf SAF}-solution because 
$u_p^2 \neq u_{\rm AW}^2$ there; such a radius is $r=r_{\rm A}^{out}$ 
for case (i) and $r=r_{\rm A}^{in}$ for case (ii). 
Outside this latter Alfv\'en radius, $B^\phi/B_p<0$ and magnetic energy 
streams outward ($nu^r |E| X_{\rm em} >0$) to the injection point, 
while inside this point $B^\phi/B_p>0$ and magnetic energy streams 
inward. The fluid part of energy flux always streams inward 
( $nu^r |E| X_{\rm fluid} < 0$ ), and it converts to electromagnetic 
energy flux as the flow falls inward. If the ideal MHD plasma streams 
near the outer light surface, $X_{\rm em}$ has a very large negative 
value, that means a very large outgoing magnetic energy flux in the 
flow. The magnetic field line is tightly wound up ($-B^\phi/B_p\gg 1$). 
To conserve the total energy flux along the accretion flow, the ingoing 
positive fluid energy flux should be also very large. When the poloidal 
motion of the plasma is very slow near the injection point, a large 
fluid energy flux must be due to the kinetic energy of the toroidal 
motion; so we can say that the origin of the large outward 
electromagnetic energy flux is the toroidal plasma motion near 
the plasma source.

\placefigure{fig:MHDaccD}

Figure~\ref{fig:MHDaccD} shows a negative energy accretion solution
($\Omega_F \tilde L>1$). We see that the Alfv\'en point locates inside 
the ergosphere (see Paper~I). The outgoing electromagnetic energy flux 
is always greater than the ingoing fluid energy flux ($X_{\rm em} < -1$ 
and $X_{\rm fluid} = -1-X_{\rm em} > 0$). The magnetic field lines are 
trailed ($B^\phi/B_p <0$) everywhere due to the black hole rotation.

For accretion with $\Omega_F \tilde L < 1$ and 
$0 < \Omega_F < \omega_{\rm H}$, the electromagnetic energy flux also 
streams outward everywhere, but at least near the event horizon the 
ingoing fluid energy flux dominates [i.e., $-1 < (X_{\rm em})_{\rm H}<0$]. 
For magnetically dominated accretion, we see that $X_{\rm em} \simeq -1$. 
We should mention that the Poynting flux passing through the event 
horizon is not modified by the plasma inertia effect, explicitly. 
This is because the toroidal magnetic field at the event horizon 
becomes $B_{\phi {\rm H}} = 
\sqrt{(-g_{\phi\phi}/\Sigma)_{\rm H}}\,(\omega_{\rm H}-\Omega_F) 
(\partial_\theta \Psi)_{\rm H}$ for any ideal MHD accretion flows, 
which is the same expression as that of the force-free case 
\citep{Znajek77}.

We should note that, in general, 
$\zeta_{\rm S} \neq \zeta_{\rm F}$ for an {\sf SAF}-solution.  
So, when we try to estimate the locations of the fast and slow 
magnetosonic points for a given $\hat\eta$, we can only obtain 
possible ranges of magnetosonic points by using the 
$r_{\rm cr}$-$\hat\eta$ diagram. For example, if we know the values 
of $r_{\rm F}$ and $\zeta_{\rm F}$, we can find a possible region 
of $r_{\rm S}$ for an acceptable $\zeta_{\rm S}$ range. 
To determine the location of the slow magnetosonic point explicitly, 
we need to obtain the $\zeta_{\rm S}$ value by solving the poloidal 
equation, which is a polynomial with high degree.

\subsection{Fluid-Dominated Flows } 

The solution for case (ii) becomes hydrodynamical accretion in 
the hydro-dominated limit. 
In the limit of weak magnetic field 
($\varepsilon\equiv B_p^2/(8\pi P) \ll 1$), 
for a flow with $u_p^2(x)\sim O(\varepsilon^{0})$, we see that 
$M^2\sim O(\varepsilon^{-1})$; 
then, we obtain  $B_\phi/B_p\sim O(\varepsilon^{0})$, 
$(E/\mu) =  u_t    +O(\varepsilon)$, and 
$(L/\mu) = -u_\phi +O(\varepsilon)$. 
The poloidal equation~({\ref{eq:pol-eq}) then becomes 
\begin{equation}
   (1+u_p^2) = -k\left(\frac{E}{\mu}\right)^2 +O(\varepsilon) \ .
\end{equation}
The function defined by equation~(\ref{eq:kk}) is reduced to 
$k = -g^{tt} +2g^{t\phi}\ell -g^{\phi\phi}\ell^2 $, where 
$\ell\equiv -u_\phi/u_t$ is the specific angular momentum as 
measured at infinity. The numerator~(\ref{eq:numer}) and 
denominator~(\ref{eq:domom}) become 
\begin{eqnarray}
  {\cal N} &=& (M^2)^3(1+u_p^2)\left[C_{\rm sw}^2  {\cal S}  
      - \frac{1}{2}(1+C_{\rm sw}^2)(\ln k)' \right] \label{eq:N-hyd}\ , \\
  {\cal D} &=& (M^2)^3(C_{\rm sw}^2-u_p^2) \ , 
\end{eqnarray}
where the function ${\cal S}={\cal S}(r,\theta)$  
is related to the configuration of a stream line; for example, for 
a radial stream line, $ {\cal S}(r) = -(2r^2-3mr+a^2)/(r\Delta) $.  
Here, we should remember that, in the ideal MHD case, a stream line 
coincides with a magnetic field line. So, even for a weak magnetic 
field, the factor ${\cal S}$ should be related to the magnetic field 
lines, and in fact it must be replaced by ${\cal S}=(\ln B_p)'$ for 
MHD flows.  
Further, we should say that, in the weak-magnetic field limit, 
$\Omega_F$ represents the angular frequency of the stream 
line for a hydrodynamical flow, and it would be determined as the 
angular velocity of the injection point. 
Thus, the above expressions are reduced to the relativistic 
hydrodynamical flow equations formulated by \citet{Lu86}.
In equations (\ref{eq:aw}), (\ref{eq:fw}) and (\ref{eq:sw}), 
we can also check that the fast magnetosonic wave speed equals 
the sound wave speed, while the Alfv\'en wave speed and the 
slow magnetosonic wave speed become zero.

To the contrary, for $u_p^2 \sim u_{\rm AW}^2 \sim O(\varepsilon)$, 
which corresponds to case (i), the hydrodynamical expression for the 
poloidal equation can not be obtained by a flow solution. At the 
Alfv\'en point, $M^2_{\rm A}=\alpha_{\rm A}\sim O(\varepsilon^{0})$ 
and $u_{\rm A}^2 \sim O(\varepsilon)$. At the fast magnetosonic point, 
$u_{\rm F}^2 \sim O(\varepsilon)$. So, the velocity of trans-Alfv\'enic 
MHD accretion would also be the order of $\varepsilon^{1/2}$.

\section{Concluding Remarks } 

We have considered stationary and axisymmetric hot ideal MHD accretion 
along a flux-tube connected from a plasma source to the event horizon.
To argue the details of the boundary conditions at the plasma source 
would take us beyond the scope of this paper. Therefore, we have 
surveyed the dependence of the trans-fast MHD flows on a wide range 
of source parameters. 
We have shown that, when the forbidden region is type IA or~IIIA, 
there exist two physically different accretion regimes: (i)
{\it magneto-like}\, MHD accretion and (ii) {\it hydro-like}\, MHD 
accretion. The magneto-like MHD accretion would be injected from near 
the separation point and passes through the inner Alfv\'en point with 
a smaller $\hat\eta$. On the other hand, the hydro-like MHD accretion 
with a sufficiently large $\hat\eta$ would be injected from between 
the outer Alfv\'en point and the outer light surface and passes through 
the outer Alfv\'en point. Hydro-like accretion may also be initially 
super-slow magnetosonic or super-Alfv\'enic. A hot ideal MHD plasma with 
larger $\hat\eta$ cannot accrete stationary onto the black hole after 
passing through the inner Alfv\'en point. 
Then, if the value of $\hat\eta$ increases with a secular timescale, 
the magneto-like MHD accretion solution should transit to the hydro-like 
MHD accretion solution; the inverse process would be also possible. 
The criterion for distinguishing between the two regimes is based on 
the locations of both the Alfv\'en point and the fast magnetosonic 
point, which change discontinuously during the transition.

For the magneto-like MHD accretion, we have found that the location of 
the X-type fast magnetosonic point is not unique for fixed intrinsic 
parameters of the accreting plasma. For example, 
in Figures~\ref{fig:eta-OM}a,~\ref{fig:eta-spin}b and~\ref{fig:eta-alf}, 
in the range of $r_{\rm H}<r_{\rm F}<r_{\rm A}^{in}$, 
a $\hat\eta=$ constant line crosses a solid curve of 
$\zeta_{\rm F}\leq 0.1$ at three points; the first and third fast 
magnetosonic points are X-type critical points, while the second one 
is O-type. So, we can expect two modes of magneto-like MHD accretion 
solutions; that is, one passes through the ``inner'' inner-fast
magnetosonic point and the other passes through the ``outer'' inner-fast 
magnetosonic point. The number of these inner-fast magnetosonic points 
depends effectively on $a$, $\delta$, $\Omega_F$, $\tilde L$, 
$\zeta_{\rm F}$ and $\hat\eta$.  
There is a tendency for larger $\Omega_F-\omega_{\rm H}$ to generate 
a multiple inner-fast magnetosonic solution. Compared with the 
radial field geometry, converging field geometries for accreting flows 
($\delta>0$) also generate such multiple inner-fast magnetosonic 
solutions. Transition between these two modes is also discontinuous. 
The $\delta$-dependence of {\sf SAF}-accretion solutions 
on the flow velocity $u^r(r)$ and the electromagnetic energy 
$X_{\rm em}(r)$ is very weak, as long as the location of the fast 
magnetosonic point does not jump, by changing the $\delta$ value, 
to another branch of the multi-inner fast magnetosonic points.  
The radial terminal velocity at the event horizon $u^r_{\rm H}$, 
rather than the radial one, slightly decreases (increases) for 
$\delta>0$ ($\delta<0$).  
For example, we have checked this for $\delta=0.4, 0.0$ and $-0.4$  
cases. For such accretion solutions, we also find that for $\delta>0$ 
($\delta<0$) the total energy flux per magnetic tube $\hat\eta E$ and 
$\hat\eta$ increase (decrease), while the total energy $E$ decreases 
(increases).

Throughout this paper, we have only discussed the ideal MHD flow cases. 
However, non-ideal MHD flow solutions near the event horizon are also 
presented (\cite{Punsly90,Punsly01}). \cite{Punsly90} discussed an 
ingoing magnetic flow solution along magnetic field lines that thread 
the ergosphere and the equatorial plane (and therefore not the event 
horizon). This solution corresponds to our sub-Alfv\'enic ingoing 
solution approaching the inner light surface with zero poloidal 
velocity (see, e.g., Fig.~\ref{fig:MHDaccA}a); along this solution 
$B^\phi=0$ at $r=r_{\rm A}^{in}$ (see also, Fig.~\ref{fig:MHDaccA}b), 
where $u_p^2\neq u_{\rm AW}^2$ (not the Alfv\'en point {\sf A}), while 
for the {\sf SAF}-solution $B^\phi\neq 0$ at the Alfv\'en point {\sf A}. 
When we consider an accretion solution onto a black hole, it seems that 
there is no sub-Alfv\'en accretion solution consistent with the ideal 
MHD approximations due to the existence of the forbidden region. 
However, of course, for such a set of ingoing flow parameters the ideal 
MHD approximation must be rejected, and then the non-ideal MHD ingoing 
flow should exist. This is because near the light surface due to the 
plasma inertia effects large radiation losses are expected, and large 
radiation losses equate to a dissipative plasma and a breakdown of the 
ideal MHD approximation. Then, a non-ideal MHD accretion flow solution, 
which does not pass through the fast-magnetosonic critical point {\sf F} 
discussed in the previous section, also exists in the region downstream 
of the light surface because of the inward attraction of black hole 
gravity.  
This requires that dissipative effects be incorporated into the physical 
description of the inward extension of the ideal MHD wind inside the light 
surface. Such a plasma propagates inside the inner light surface and 
enters the forbidden region of $r\leq r_{\rm L}^{in}$ with relatively 
slow velocity (as compared with the Alfv\'en wave speed except for the 
area close to the horizon) by crossing or reconnecting the magnetic  
field lines, where the physical meaning of the concept of the forbidden 
region for the ideal MHD flows would be lost. Note that the non-ideal 
MHD accretion flow must also become super-Alfv\'en and super-fast 
magnetosonic just inside the inner light surface (see Chapter~9 
of Punsly 2001).

For a ``super-critical'' accretion flow of $E>E_{\rm F}$ and 
$u_{\rm AW}^2 < u_p^2 < u_{\rm FM}^2$, which reaches the event horizon 
without passing through the fast magnetosonic point {\sf F} and is 
unphysical under the ideal MHD approximation, some kinds of dissipative 
effects near the event horizon would also make the super-critical 
accretion onto the black hole possible.    
One may also expect a ``sub-critical'' accretion flow of $E<E_{\rm F}$,  
which is also not a solution for the inflow into the black hole under 
the ideal MHD approximation. However, the breakdown of the ideal MHD 
would change the nature of the critical points (see, e.g., 
\cite{Chakrabarti90} for the hydrodynamical flow case); that is, instead 
of the X-type and O-type critical points, the ``nodal-type'' and 
``spiral-type'' critical points appear on the flow solutions in the 
($r, u^r$)-plane. 
Thus, when the initial condition at the plasma source does not match 
the critical condition at the fast magnetosonic point, the non-ideal 
MHD inflow into the black hole must be realized. The detailed structure 
of the non-ideal MHD flows around the critical point is a very important 
topic for black hole accretion, but further discussion is out of the 
scope of this paper.

When the angular velocity of the magnetosphere is in the range of 
$0<\Omega_F<\omega_{\rm H}$, only one Alfv\'en point appears on the 
$r$-$u^r$ plane. To accrete onto the black hole, an ideal MHD flow 
must pass through the Alfv\'en point classified by type IIA$^{in}$,  
IIA$^{out}$, IIB$^{in}$ or IIB$^{in}_\ast$ in Paper~I. Determining whether 
the accretion is magneto-like or hydro-like may not be clear in type 
IIA case. However, we can expect that the Alfv\'en point {\sf A}$^{in}$ 
is responsible for the magneto-like accretion solution and the 
{\sf A}$^{out}$ point is responsible for the hydro-like accretion 
solution. It seems that for the Alfv\'en point of type IIB$^{out}$ 
there are no accretion solutions consistent with the ideal MHD  
approximation.     
Of course, for such a set of ingoing flow parameters the ideal MHD 
approximation must be abandoned. 
When the {\sf SAF}-accretion solution is invalid 
(e.g., the type~IIB$^{out}$ case), the non-ideal MHD accretion becomes 
important and would dynamically effect the magnetic field structure.

When a region of $k(r) > 0$ exists along a magnetic field line for 
larger $\tilde L\Omega_F \, (\leq 1)$, the forbidden region is type~B 
and the ideal MHD accretion is only allowed after passing the 
{\it inner}\/ Alfv\'en point. The hydro-like MHD accretion does not 
arise because of the sufficiently strong centrifugal barrier, while 
the magneto-like MHD accretion is available because of the effective 
angular momentum transport from the fluid-part of total angular momentum
to the magnetic-part (Hirotani et~al.~1992). When the hot effects 
dominate in the plasma, this magneto-like MHD accretion may be also 
forbidden due to the disappearance of the inner-fast magnetosonic 
point. In this case, however, we can also expect the non-ideal MHD 
accretion to fall into the black hole.

If a shock front is generated after passing the fast magnetosonic 
point, the post-shock flow with increased entropy must pass another 
fast magnetosonic point again on the way to the event horizon. 
To construct such a shock formation model, the existence of multiple 
fast magnetosonic points is required in the accretion solution. We can 
expect two types of discontinuous transitions. One is the transition 
from the hydro-like solution to the magneto-like solution at somewhere 
between the middle-fast and inner-fast magnetosonic points. The other 
is the transition in the hydro-like or magneto-like solution. For the 
magneto-like solution, we can see the possibility of transition from 
the $r_{\rm cr}$-$\hat\eta$ diagrams. That is, the accreting matter 
passes through the first inner-fast magnetosonic point located just 
inside the inner-Alfv\'en point, and then after the shock it passes 
through the third (X-type) inner-fast magnetosonic point. Here, the 
first and third inner-fast magnetosonic points are located on different 
$\zeta_{\rm F}$-curves in the $r_{\rm cr}$-$\hat\eta$ diagram because 
of the entropy generation. The insights gained in the course of our 
analysis should be of use in further investigations of shocked accretion 
solutions.

Although we have discussed accretion flows onto a black hole, 
our treatment can be applied to outgoing flows (i.e., winds and jets). 
To do so, we can plot $r_{\rm cr}$-$\hat\eta$ diagrams in 
the range of $r_{\rm L}^{in} < r_{\rm cr} < \infty$ (for the fast 
magnetosonic point, $r_{\rm A}^{in} < r_{\rm F} < \infty$; for the slow 
magnetosonic point, $r_{\rm L}^{in} < r_{\rm S} < r_{\rm A}^{out}$). 
So, we will find the possible locations of the magnetosonic point for 
outflows. When the forbidden region is type IA or IIIA, we find two 
regimes of {\sf SAF}-solutions for outflows; that is, ${\sf inj}$ 
$\to$ ${\sf S}^{\rm mid}$ $\to$ ${\sf A}^{out}$ $\to$ 
${\sf F}^{\rm out}$ $\to$ $\infty$ and ${\sf inj}$ $\to$ 
${\sf S}^{\rm in}$ $\to $ ${\sf A}^{in}$ $\to$ ${\sf F}^{\rm mid}$ 
$\to$ $\infty$, where the slow magnetosonic point ${\sf S}^{\rm in}$ 
is located between the inner light surface and the inner Alfv\'en point, 
and the fast magnetosonic point ${\sf F}^{\rm out}$ is located outside 
the outer Alfv\'en point.  
The former {\sf SAF}-solution is also available for the type B forbidden 
region, and the latter {\sf SAF}-solution remains in the hydrodynamical 
limit.  
From the $r_{\rm cr}$-$\hat\eta$ diagrams, for the outflow, the location 
of the fast magnetosonic points with the same $\hat\eta$ value strongly 
depends on $\delta$. A similar result has been discussed in 
\citet{Takahashi-Shibata98} as a pulsar wind model without the 
gravitational effects. Here, we should note that to blow away to 
infinity, the outgoing flows also must pass through the slow 
magnetosonic point and the Alfv\'en point, but at the asymptotic region 
both the super-fast and sub-fast magnetosonic outflows are available. 
Whether or not the outflow passes through the fast magnetosonic point 
depends on the field aligned parameters of flows and the geometry of 
field lines.  
So, for the outflow, the fast-magnetosonic surface does not need to 
cover all solid angles. We can expect that the trans-fast MHD outflow 
is realized at least in some part of the magnetic field lines, and the 
distribution of the fast-magnetosonic surface would play a very 
important part in explaining the generation and collimation of a highly 
accelerated jet or wind.

\acknowledgments

We are grateful to Akira Tomimatsu, Sachiko Tsuruta  and 
Vasily S. Beskin for useful discussions and to an anonymous 
referee for constitutive criticism which helped us to improve the paper.

\clearpage 

\clearpage


\begin{figure}
    \epsscale{0.7}
    \plotone{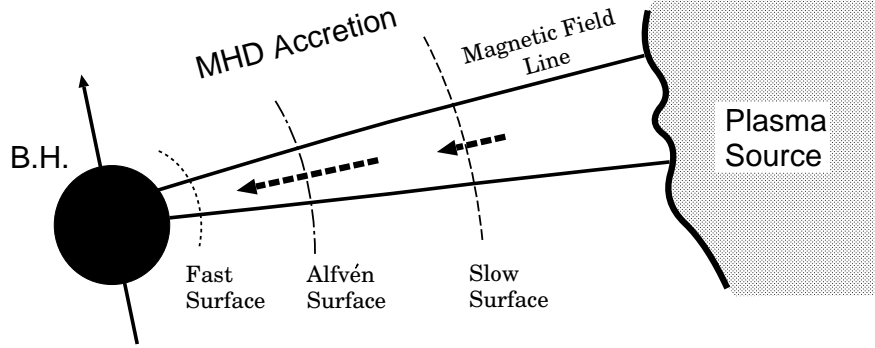} 
\caption{
         MHD accretion onto a black hole. The flow injected 
         from a plasma source (e.g., the disk surface or the corona) 
         passes through the slow magnetosonic point, the Alfv\'en 
         point and the fast magnetosonic point, in order, and then 
         goes across the event horizon.   
        }
\label{fig:acc}
\end{figure} 

\begin{figure}[t]
  \epsscale{1.0}
  \plottwo{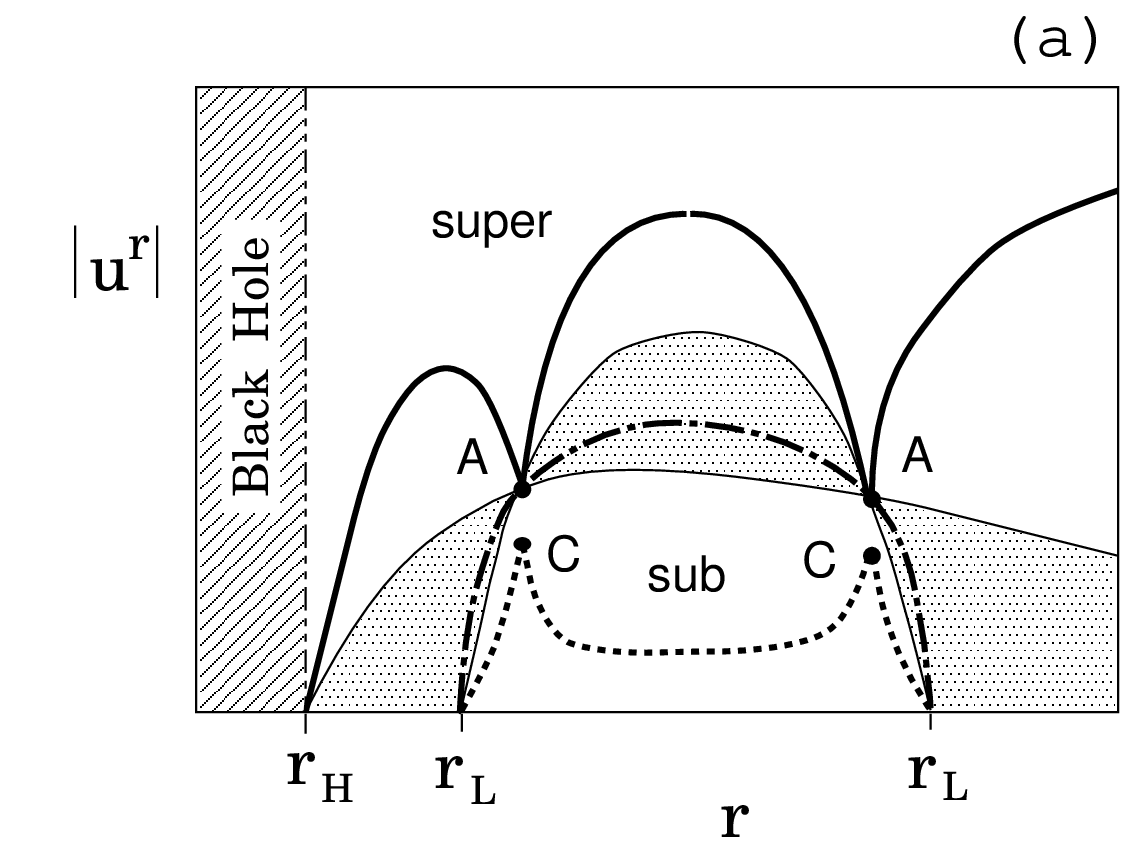}{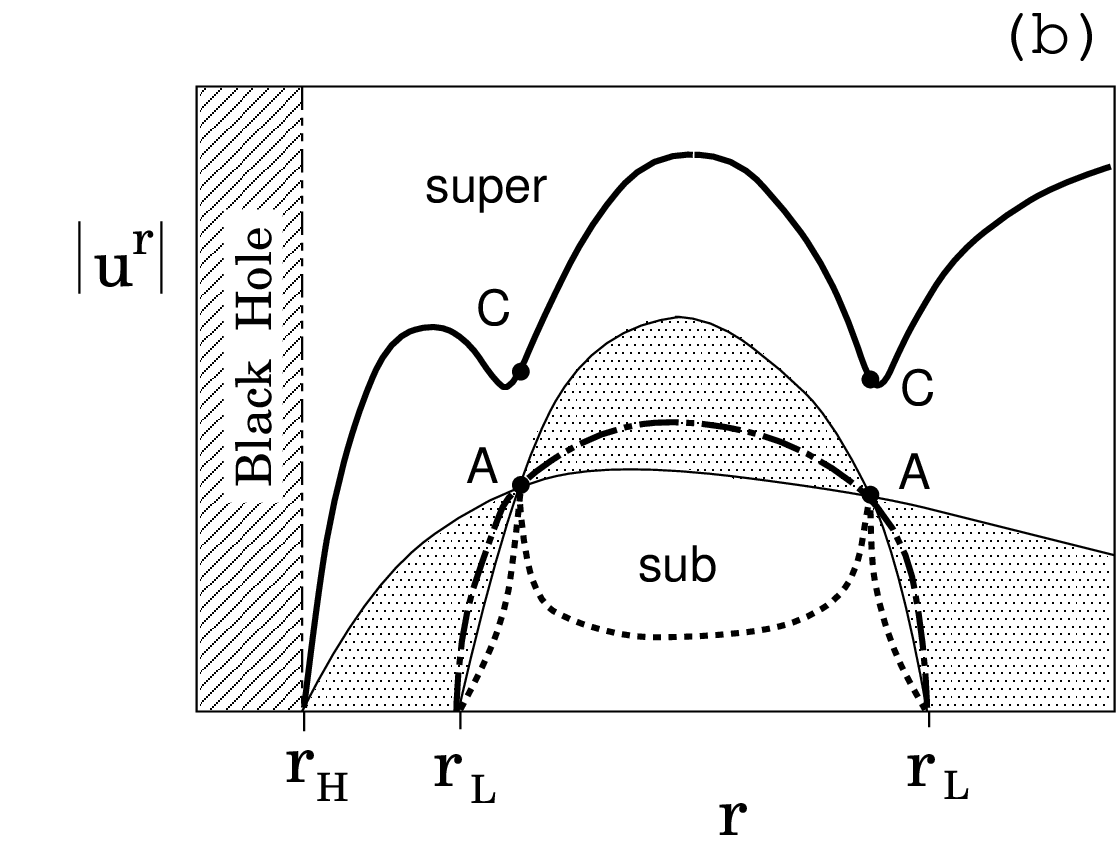} 
  \caption{
  Schematic pictures of ${\cal D}(r, u^r)=0$ curves for  
  (a) strong magnetic field $(C_{\rm sw}^{2})_{\rm A} < u_{\rm A}^2$ 
  and   (b) weak magnetic field $(C_{\rm sw}^{2})_{\rm A} > u_{\rm A}^2$. 
  The thick-solid-curves correspond to the $u_p^2 = u_{\rm FM}^2$
  curves; the thick-dash-dot-curves correspond to the 
  $u_p^2 = u_{\rm AW}^2$ curves; and the thick-dotted-curves correspond 
  to the $u_p^2 = u_{\rm SM}^2$ curves. The shaded regions are forbidden 
  regions of flows (type~IA or~IIIA), and separate the super-Alfv\'enic
  region labeled by ``{\sf super}'' from the sub-Alfv\'enic region 
  labeled by ``{\sf sub}''.  The hatched region is inside the black hole. 
   } 
\label{fig:dd}
\end{figure} 

\begin{figure}
  \epsscale{0.5}
  \plotone{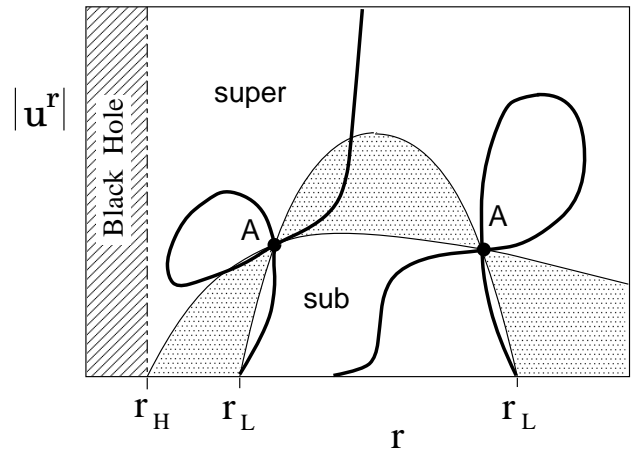}  
  \caption{
A schematic picture of ${\cal N}(r, u^r)=0$ curves. 
The shaded regions are forbidden regions (type~IA or~IIIA). 
   } 
\label{fig:nn}
\end{figure} 

\begin{figure}
   \epsscale{0.6}
  \plotone{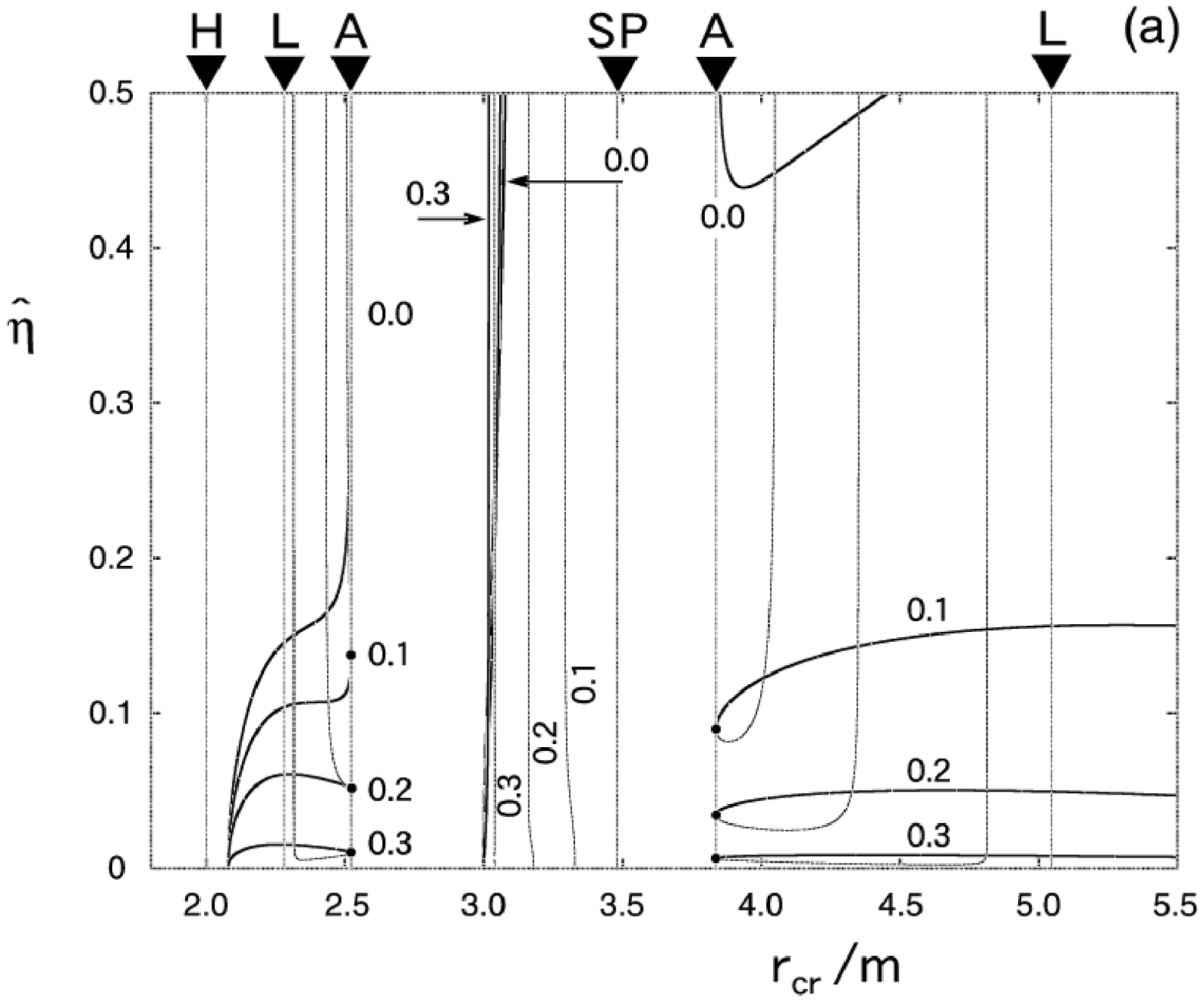}
  \plotone{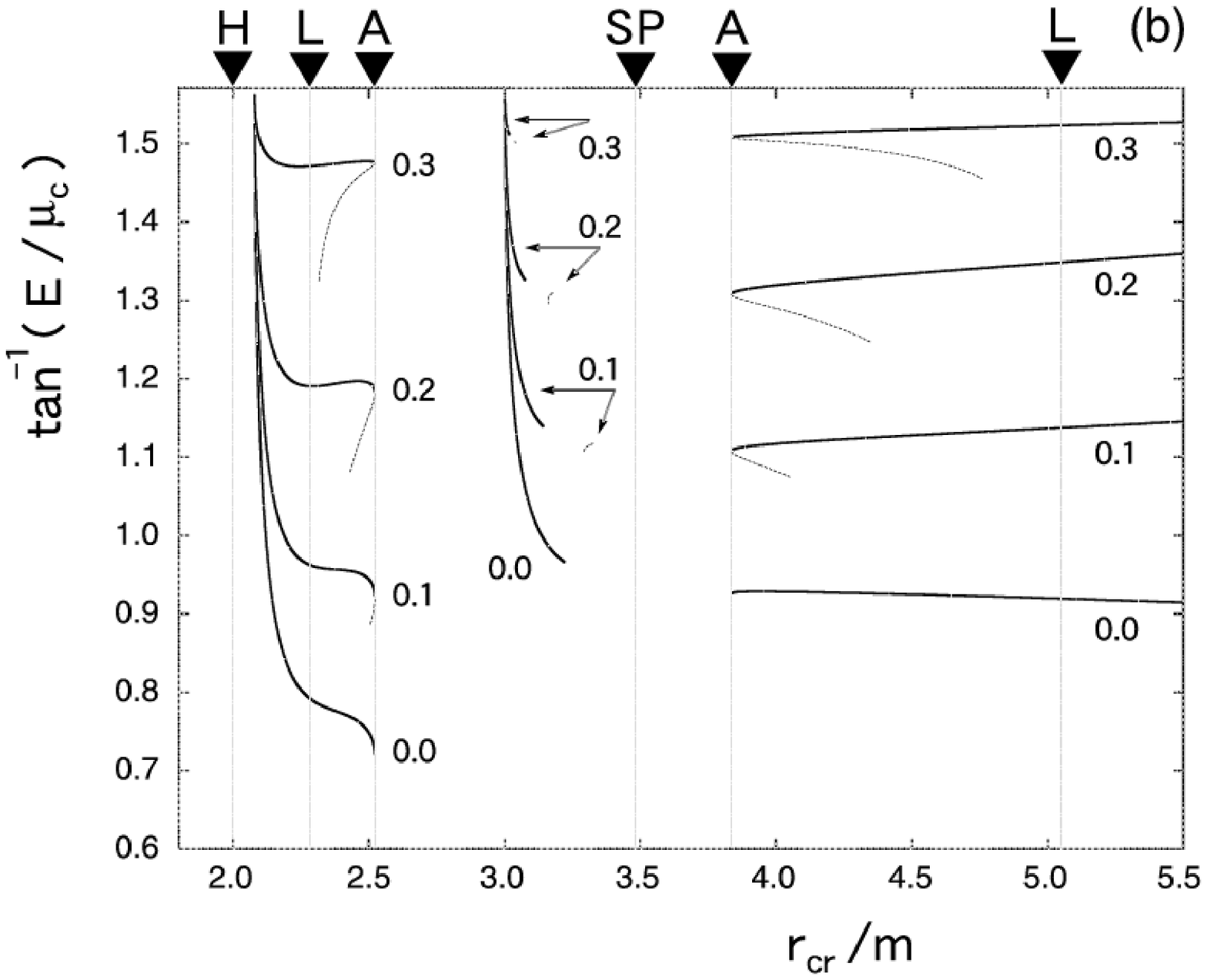}
  \caption{ 
        (a) Relations between $r_{\rm cr}$ and $\hat\eta$ and 
        (b) relations between $E/\mu_{\rm c}$ and $r_{\rm cr}$ 
        for a radial flow  with $\delta=0$ 
        in a Schwarzschild black hole magnetosphere 
        with several $\zeta_{\rm cr}$ values 
       (0.00, 0.10, 0.20, 0.30). 
        The angular velocity of the magnetic field is 
        $\Omega_F=0.8\Omega_{\rm max}$, where 
        $m\Omega_{\rm max}=0.192$. 
        The location of the Alfv\'en point is specified by 
        $x_{\rm A}=0.8$, which corresponds to 
        $\Omega_F \tilde L = 0.7287$. 
        The solid-curves indicate 
        $\hat\eta=\hat\eta(r_{\rm F})$ with $r_{\rm cr}=r_{\rm F}$.   
        The dashed-curves indicate $\hat\eta=\hat\eta(r_{\rm S})$ 
        with $r_{\rm cr}=r_{\rm S}$. 
        The shape of forbidden regions corresponds to type~IA. 
        } 
  \label{fig:eta}
\end{figure} 

\begin{figure}
  \epsscale{0.6}
   \plotone{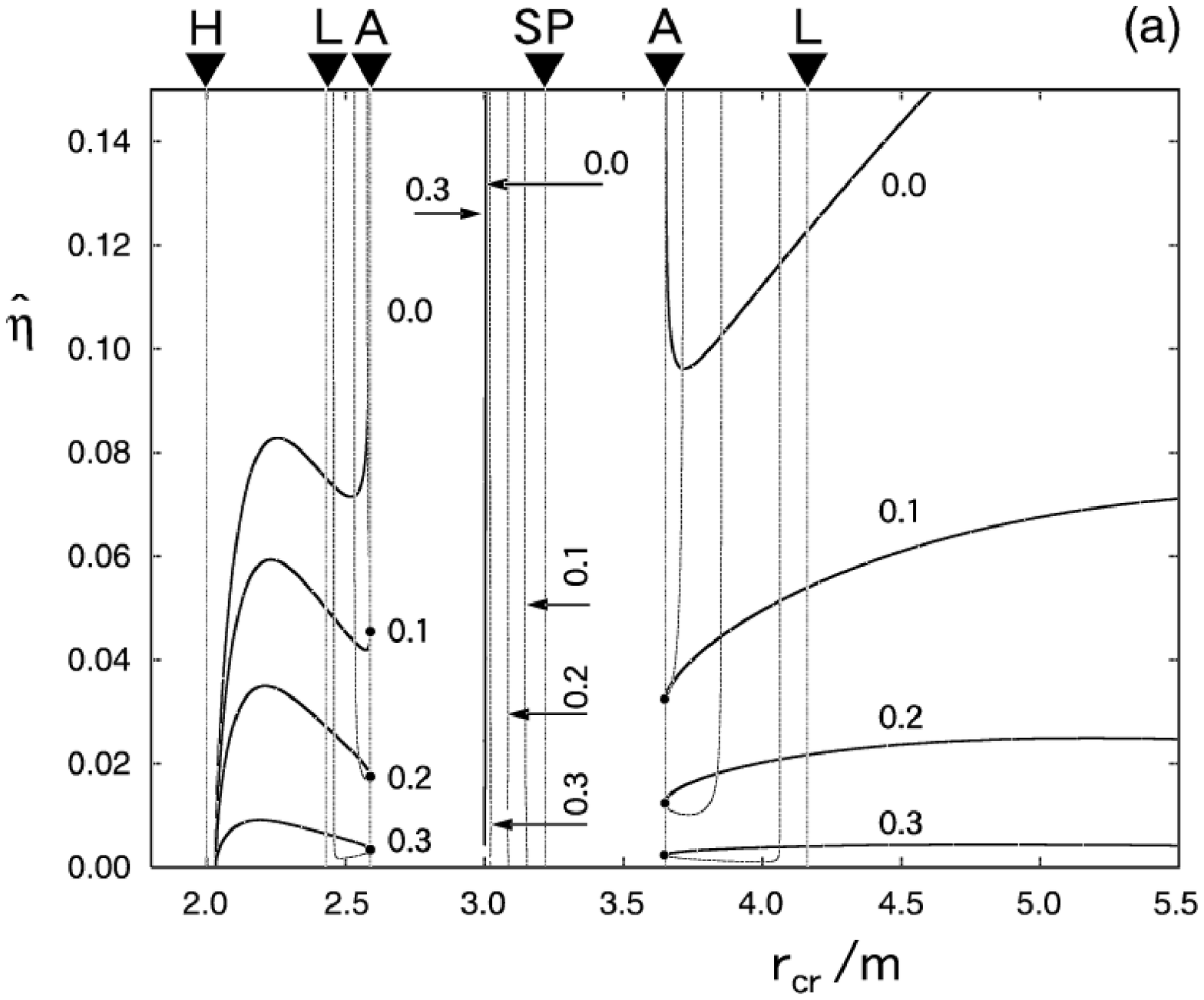}
   \plotone{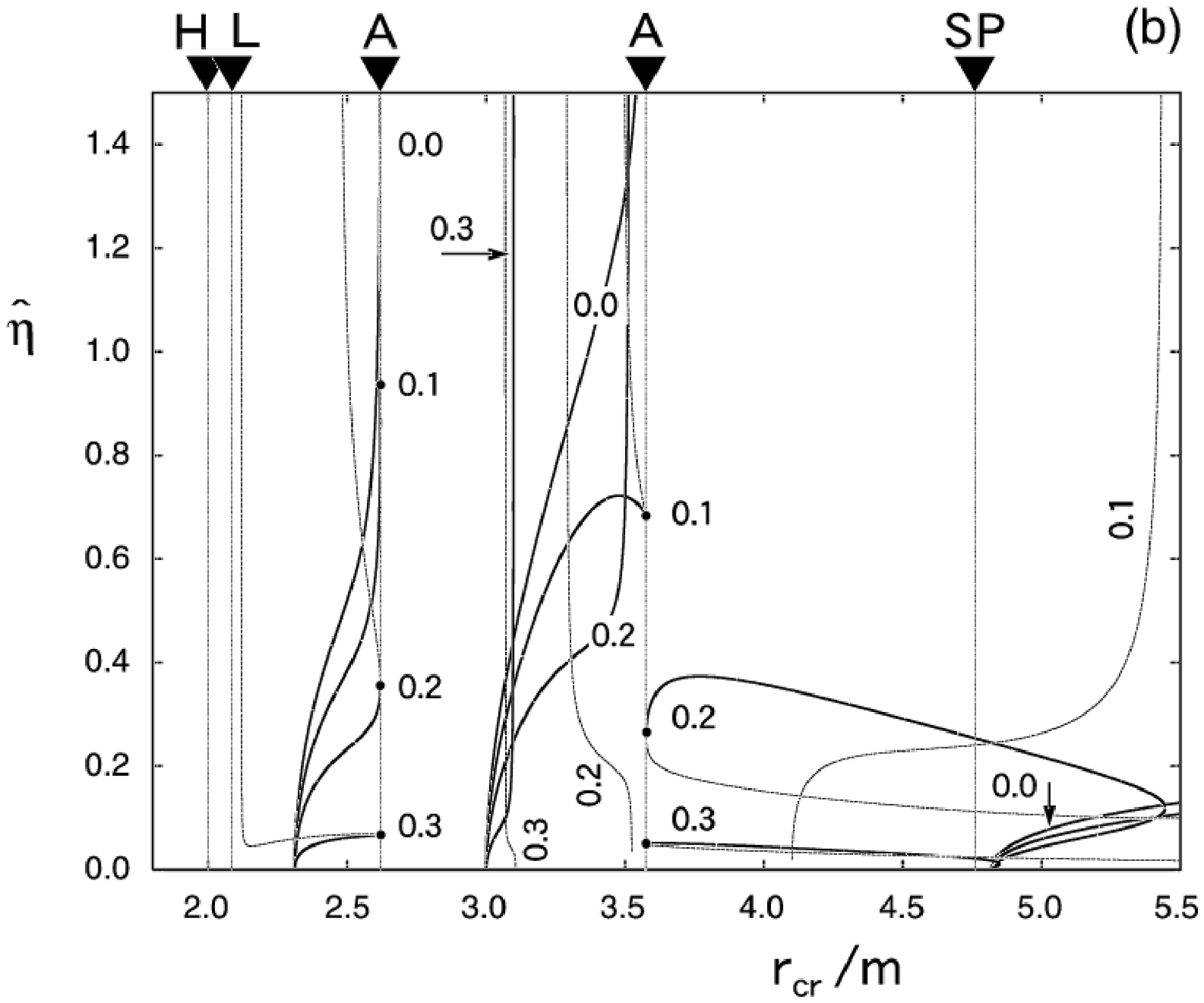}
  \caption{ 
        Relations between $\hat\eta$ and $r_{\rm cr}$ for a radial flow 
        with $\delta=0$ in a Schwarzschild black hole magnetosphere 
        with several $\zeta_{\rm cr}$ values 
       (0.00, 0.10, 0.20, 0.30). 
       The angular velocities of the magnetic field are
       (a) $\Omega_F=0.9\Omega_{\rm max}$ and 
       (b) $\Omega_F=0.5\Omega_{\rm max}$.  
       The location of the Alfv\'en point is specified by 
       $x_{\rm A}=0.8$, which corresponds to 
       (a) $\Omega_F \tilde L = 0.8841$ and 
       (b) $\Omega_F \tilde L = 0.2687$. 
       The solid-curves indicate $\hat\eta=\hat\eta(r_{\rm F})$, and 
       the dashed-curves indicate $\hat\eta=\hat\eta(r_{\rm S})$. 
       The shape of the forbidden regions corresponds to type~IA. 
        } 
  \label{fig:eta-OM}
\end{figure} 

\begin{figure}
  \epsscale{0.6}
  \plotone{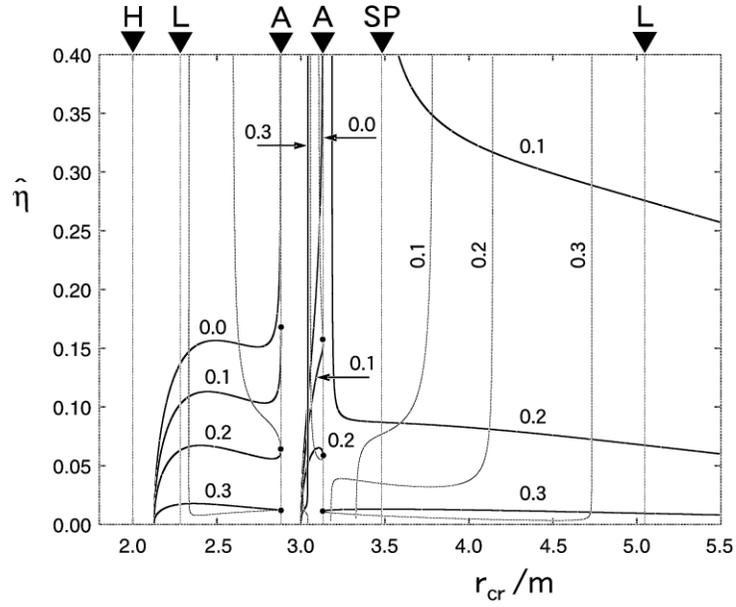}
  \caption{ 
        Relations between $\hat\eta$ and $r_{\rm cr}$ for a radial flow 
        with $\delta=0$ and several $\zeta_{\rm cr}$ values 
        (0.00, 0.10, 0.20, 0.30) in a Schwarzschild black hole 
        magnetosphere. The angular velocity of the magnetic field is 
        $\Omega_F=0.8\Omega_{\rm max}$. 
        The location of the Alfv\'en point is specified by 
        $x_{\rm A}=0.5$, which corresponds to $\Omega_F \tilde L = 0.6434$. 
        The solid-curves indicate $\hat\eta=\hat\eta(r_{\rm F})$, and 
        the dashed-curves indicate $\hat\eta=\hat\eta(r_{\rm S})$. 
        The shape of the forbidden regions corresponds to type~IA. 
        } 
  \label{fig:eta-alf}
\end{figure} 

\begin{figure}
   \epsscale{0.6}
   \plotone{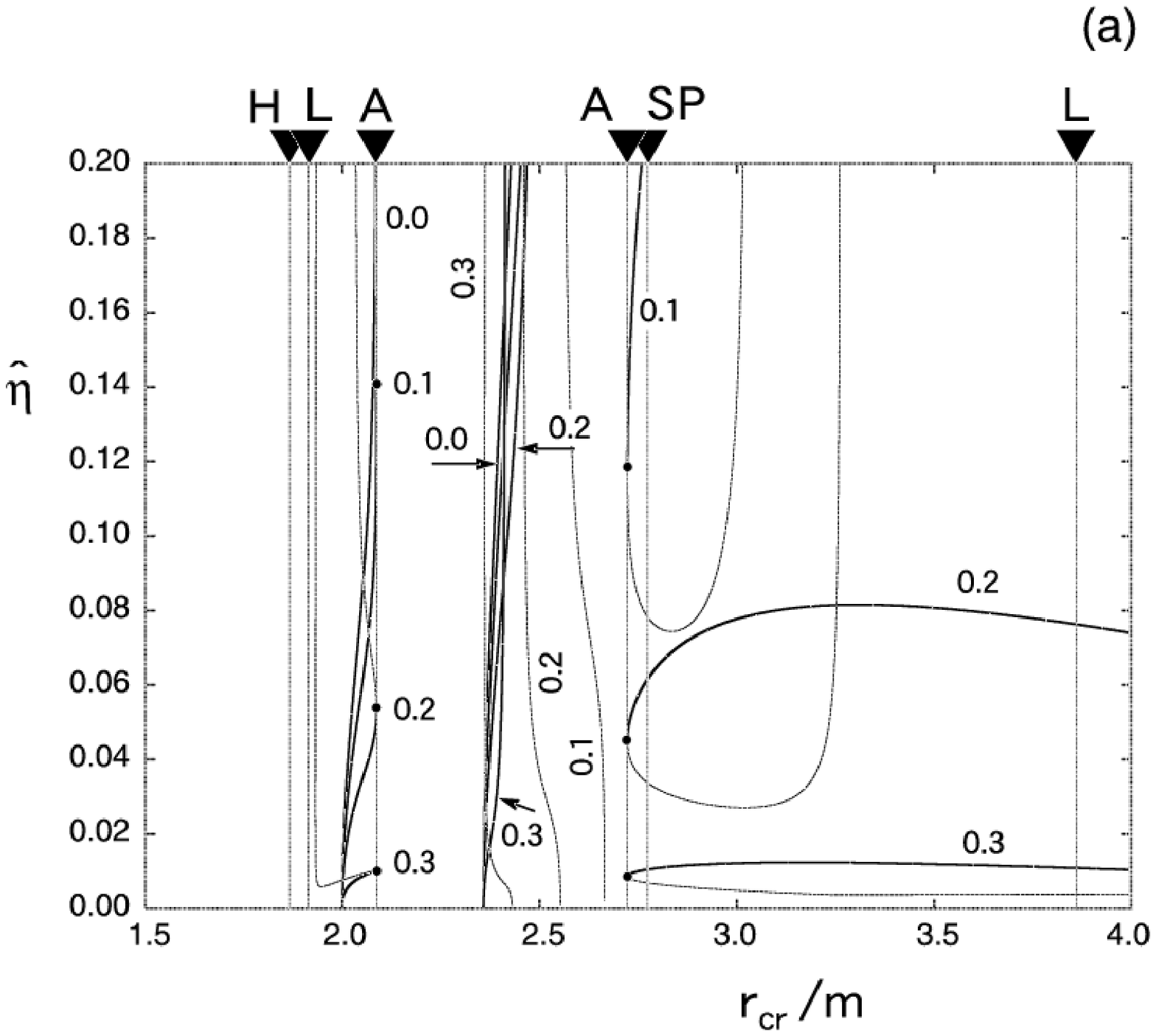}
   \plotone{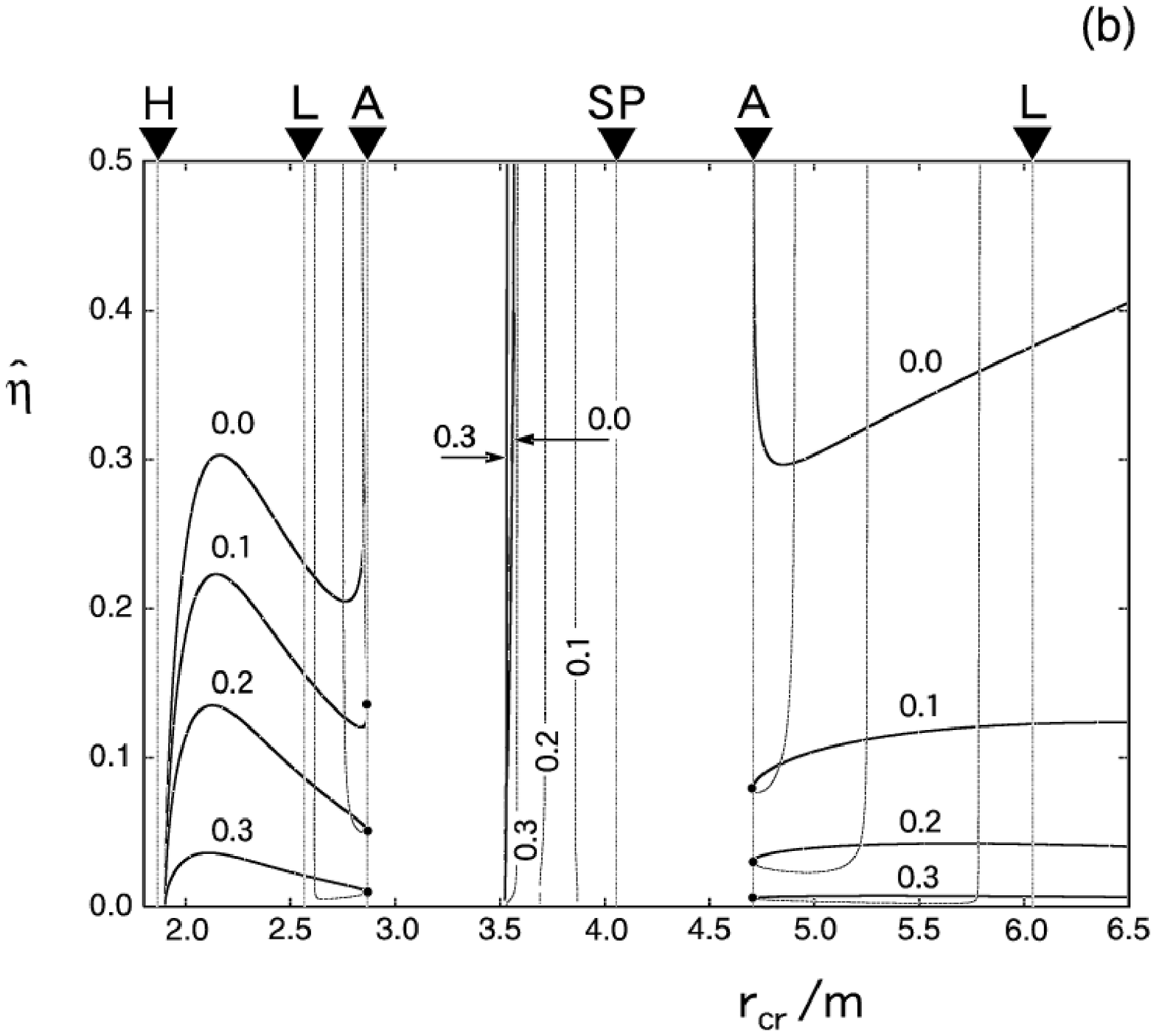}
  \caption{ 
       Relations between $\hat\eta$ and $r_{\rm cr}$ for a radial flow 
       with $\delta=0$ in a Kerr black hole magnetosphere;  
       (a) $a=0.5m$ and (b) $a=-0.5m$ with several 
       $\zeta_{\rm cr}$ values (0.00, 0.10, 0.20, 0.30). 
       The angular velocity of the magnetic field is 
       $\Omega_F=0.8\Omega_{\rm max}$, where 
       (a) $m\Omega_{\rm max}=0.244$ and (b) $m\Omega_{\rm max}=0.163$.  
       The location of the Alfv\'en point is specified by 
       $x_{\rm A}=0.8$, which corresponds to 
       (a) $\Omega_F \tilde L = 0.6749$ and 
       (b) $\Omega_F \tilde L = 0.7490$. 
       The solid-curves indicate $\hat\eta=\hat\eta(r_{\rm F})$, and 
       the dashed-curves indicate $\hat\eta=\hat\eta(r_{\rm S})$. 
       The shape of the forbidden regions corresponds to type~IA. 
        } 
  \label{fig:eta-spin}
\end{figure} 

\begin{figure}
   \epsscale{0.6}
   \plotone{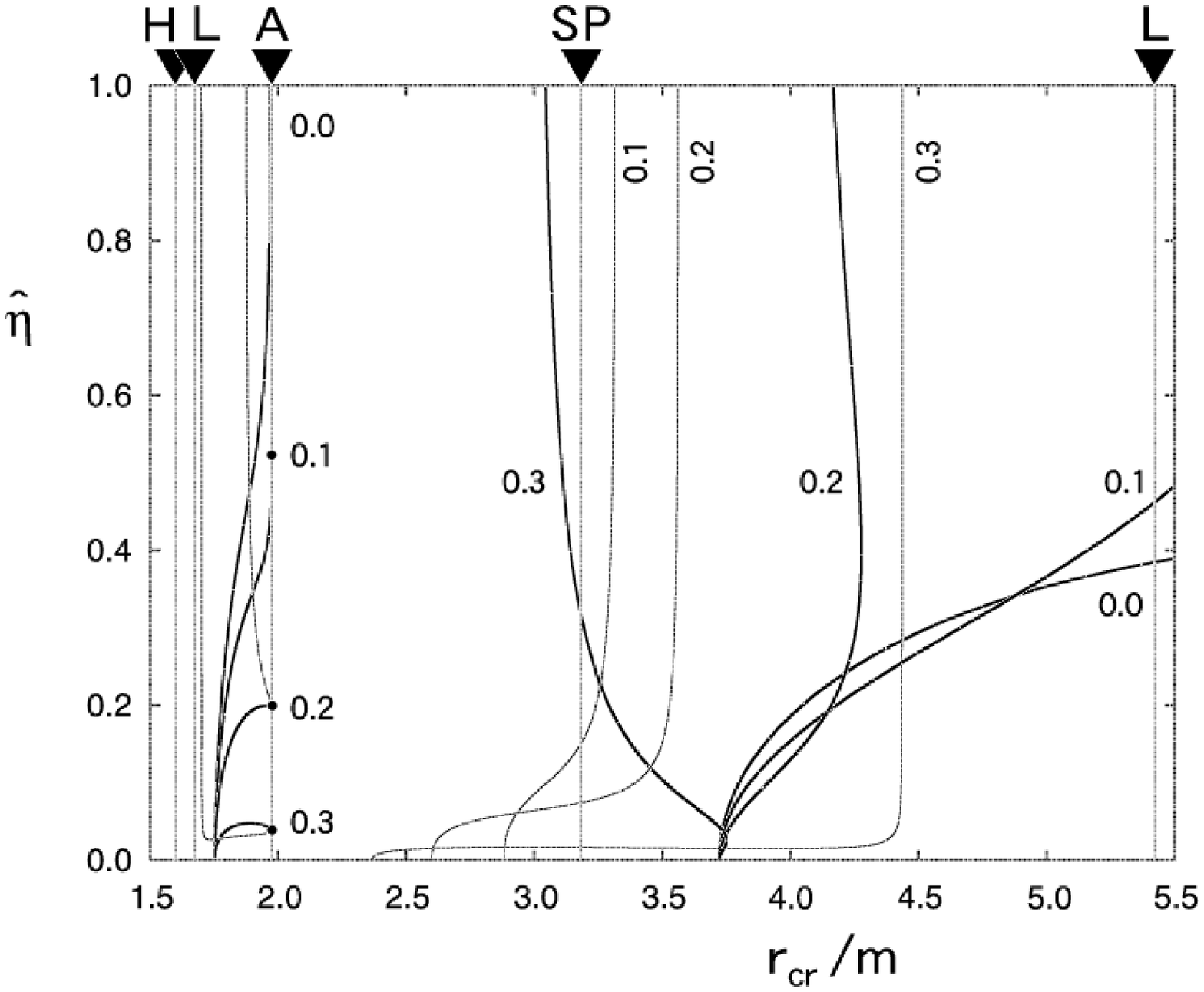}
  \caption{ 
        Relations between $\hat\eta$ and $r_{\rm cr}$ for a radial flow 
        with $\delta=0$ and several $\zeta_{\rm cr}$ values 
        (0.00, 0.10, 0.20, 0.30) in a Kerr black hole magnetosphere 
        ($a=0.8m$).  The angular velocity of the magnetic field is 
        $\Omega_F=0.5\Omega_{\rm max}$, where $m\Omega_{\rm max}=0.309$. 
        The location of the Alfv\'en point is specified by 
        $x_{\rm A}=0.8$, which corresponds to $\Omega_F \tilde L =-0.0117$.  
       The solid-curves indicate $\hat\eta=\hat\eta(r_{\rm F})$, and 
       the dashed-curves indicate $\hat\eta=\hat\eta(r_{\rm S})$. 
       The shape of the forbidden regions corresponds to type~IIA. 
        } 
  \label{fig:eta-typeII}
\end{figure} 

\begin{figure}
  \epsscale{0.6}
  \plotone{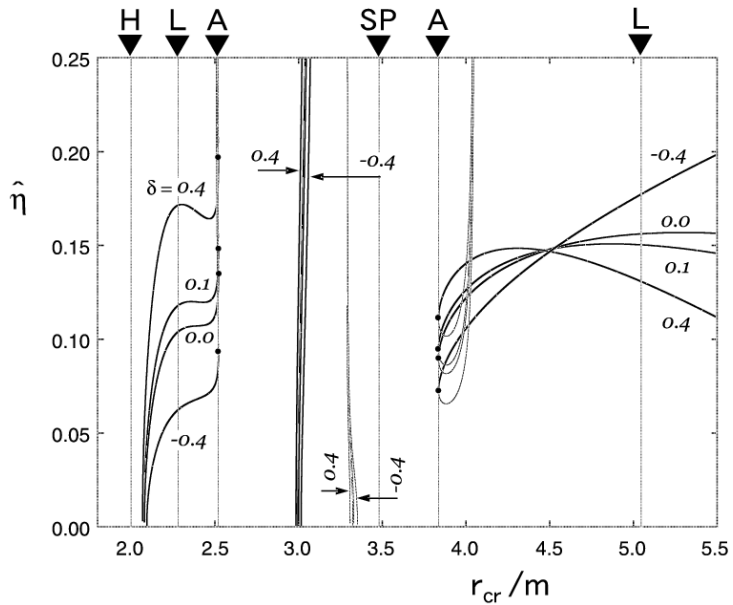}
  \caption{ 
       The $\delta$-dependence ($\delta=-0.4, 0.0, 0.1, 0.4$) 
       of the relations between $\hat\eta$ and $r_{\rm cr}$, 
       where $a=0.0$, $\zeta_{\rm cr}=0.1$, 
       $\Omega_F = 0.8 \Omega_{\rm max}$ and $x_{\rm A}=0.8$ 
       (see also Fig.~\ref{fig:eta}a).
       The solid-curves indicate $\hat\eta=\hat\eta(r_{\rm F})$, and 
       the dashed-curves indicate $\hat\eta=\hat\eta(r_{\rm S})$. 
       The shape of the forbidden regions corresponds to type~IA. 
        } 
  \label{fig:delta}
\end{figure} 

\begin{figure}
  \epsscale{0.75}
  \plotone{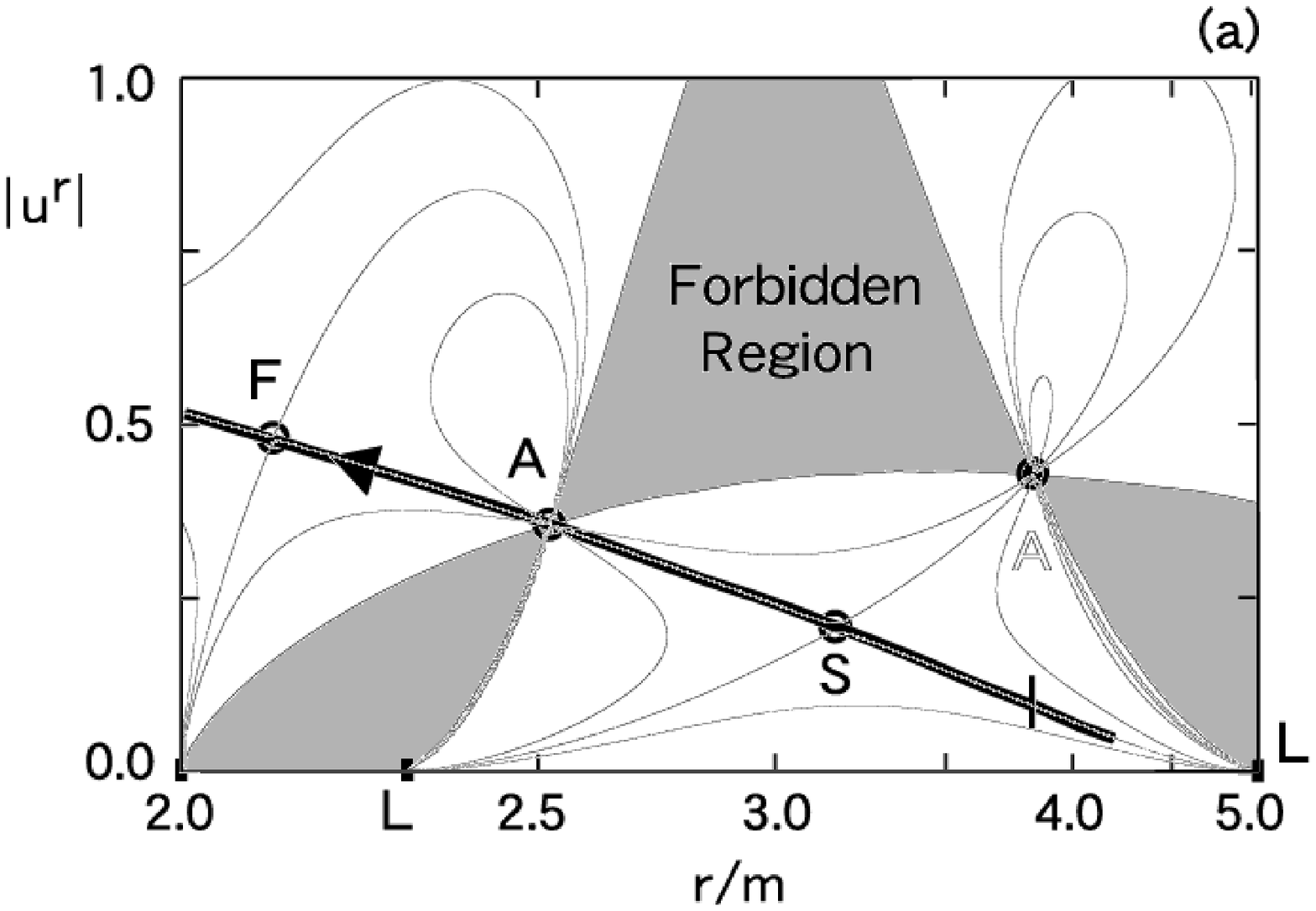} 
  \plotone{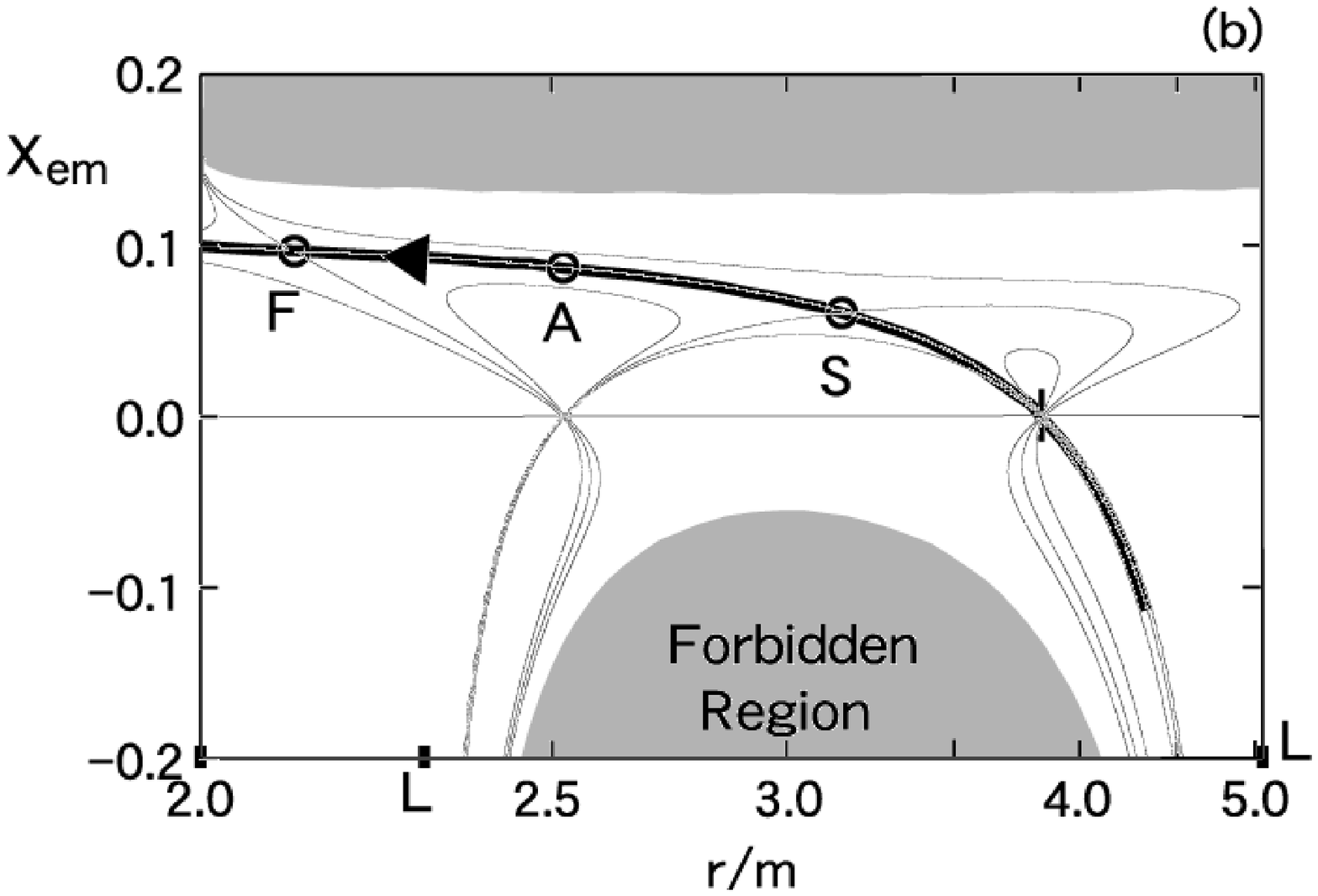} 
\caption{ 
  An example of the trans-fast MHD accretion solution (thick curves); 
  (a) radial 4-velocity and (b) the ratio of the electromagnetic energy 
  to the total energy. This solution passes through the inner Alfv\'en  
  point and the inner fast magnetosonic point ({\sf S} $\to$ {\sf A} 
  $\to$ {\sf F} $\to$ {\sf H}), where $a=0.0$, 
  $\Omega_F = 0.8\Omega_{\rm max}$, $\delta=0.0$, $\Gamma=4/3$ and 
  $\zeta_{\rm F}=0.2$. 
  The location of the Alfv\'en point is specified by $x_{\rm A}=0.8$, 
  which corresponds to $\Omega_F \tilde L=0.7287$. 
  The location of the fast magnetosonic point is  $r_{\rm F}=2.10935m$, 
  which gives $E_{\rm F}/\mu_{\rm c}= 3.6643$ and $\hat\eta=0.0337$.  
  The requirement that $\hat\eta < \hat\eta_{\rm max}$ is satisfied.  
        } 
\label{fig:MHDaccA}
\end{figure} 

\begin{figure}
   \epsscale{0.75}
   \plotone{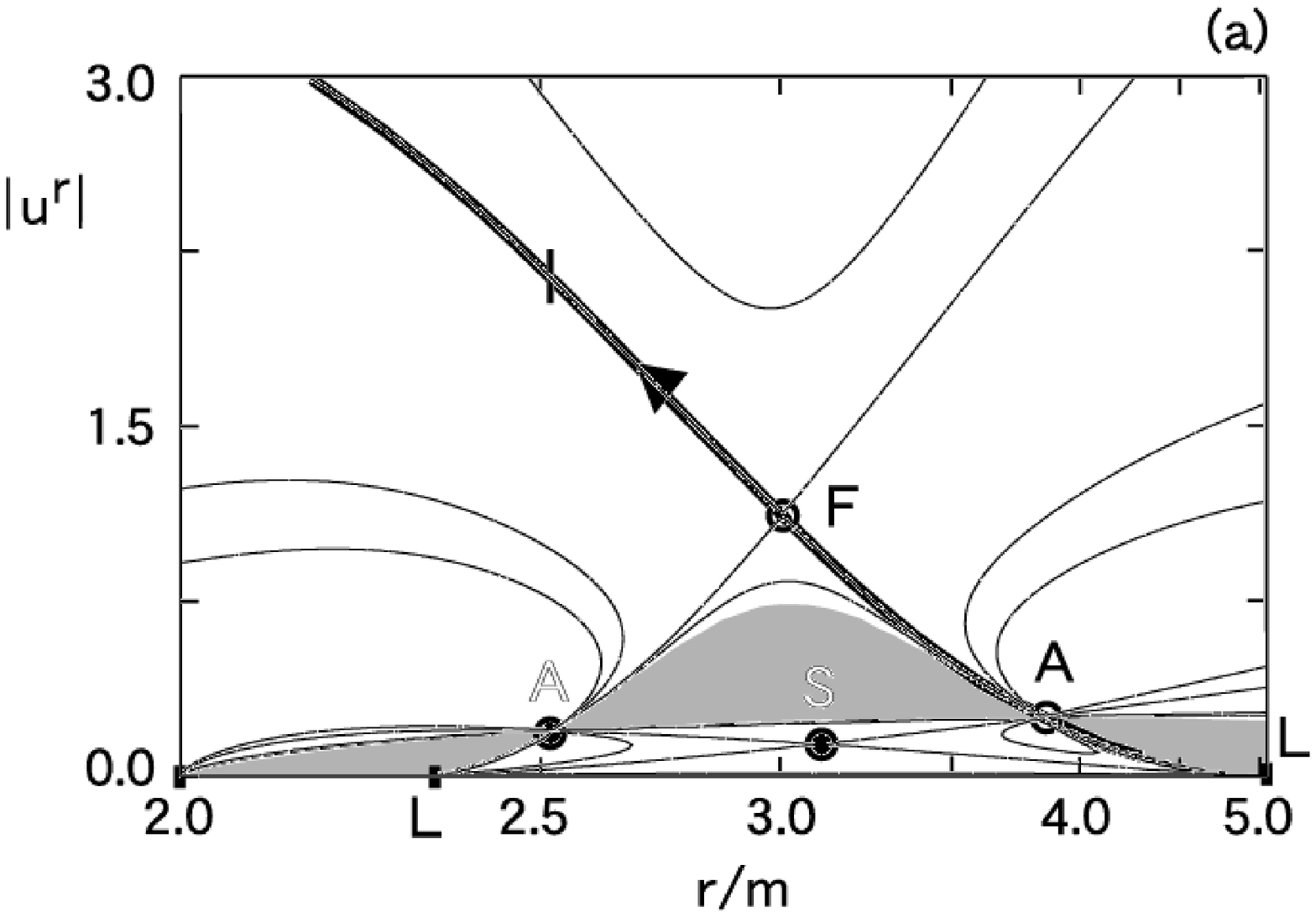} 
   \plotone{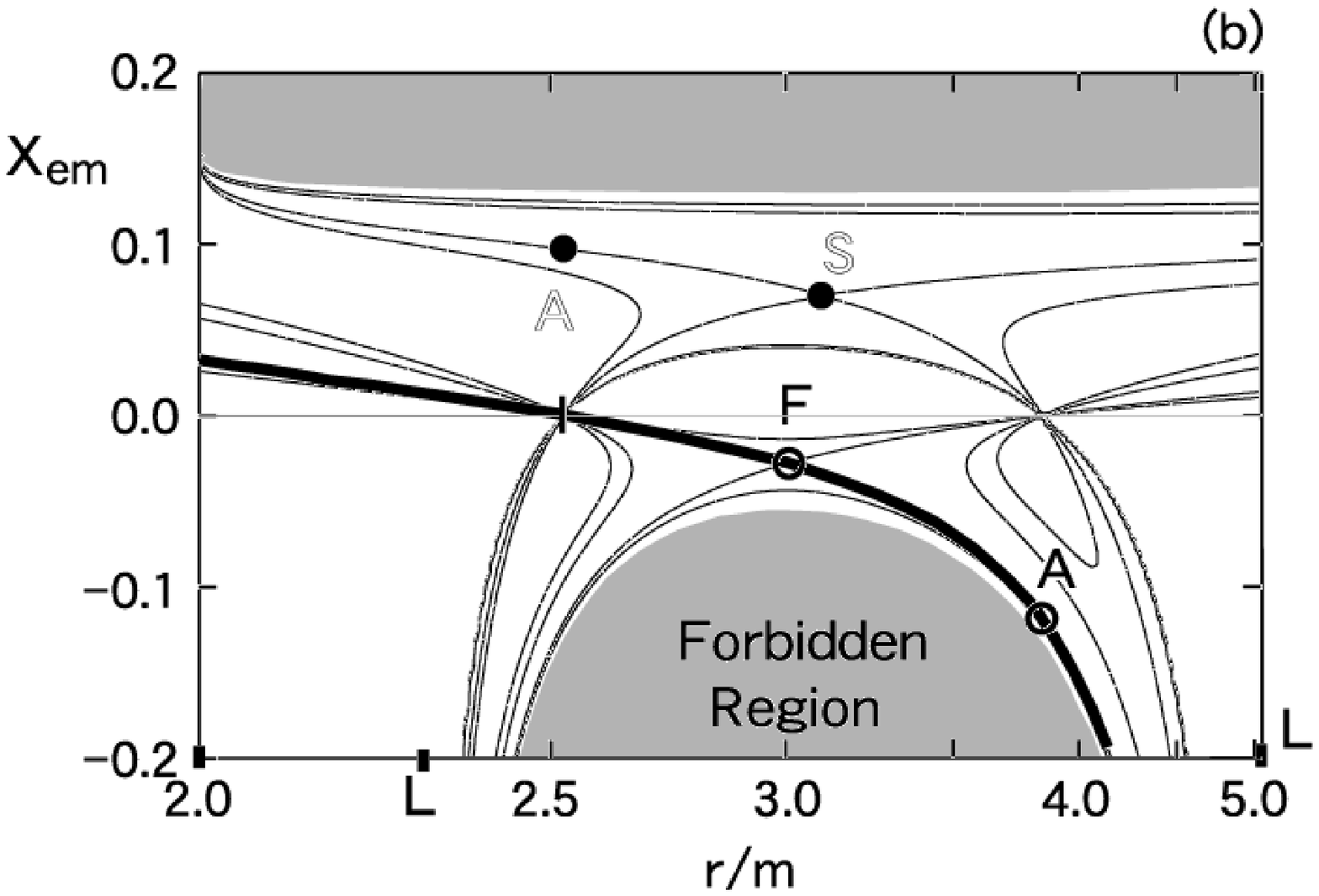} 
\caption{ 
  An example of the trans-fast MHD accretion solution (thick curves) 
  passing through the outer Alfv\'en point and the middle-fast 
  magnetosonic point; (a) radial 4-velocity and (b) the ratio of the 
  electromagnetic energy to the total energy.
  The location of the fast magnetosonic point is $r_{\rm F}=3.00490m$, 
  which gives $E_{\rm F}/\mu_{\rm c}= 9.8666$ and 
  $\hat\eta=0.0378$.  
  The other parameters are the same as in Fig.~\ref{fig:MHDaccA}. 
        } 
\label{fig:MHDaccB}
\end{figure} 

\begin{figure}
   \epsscale{0.75}
   \plotone{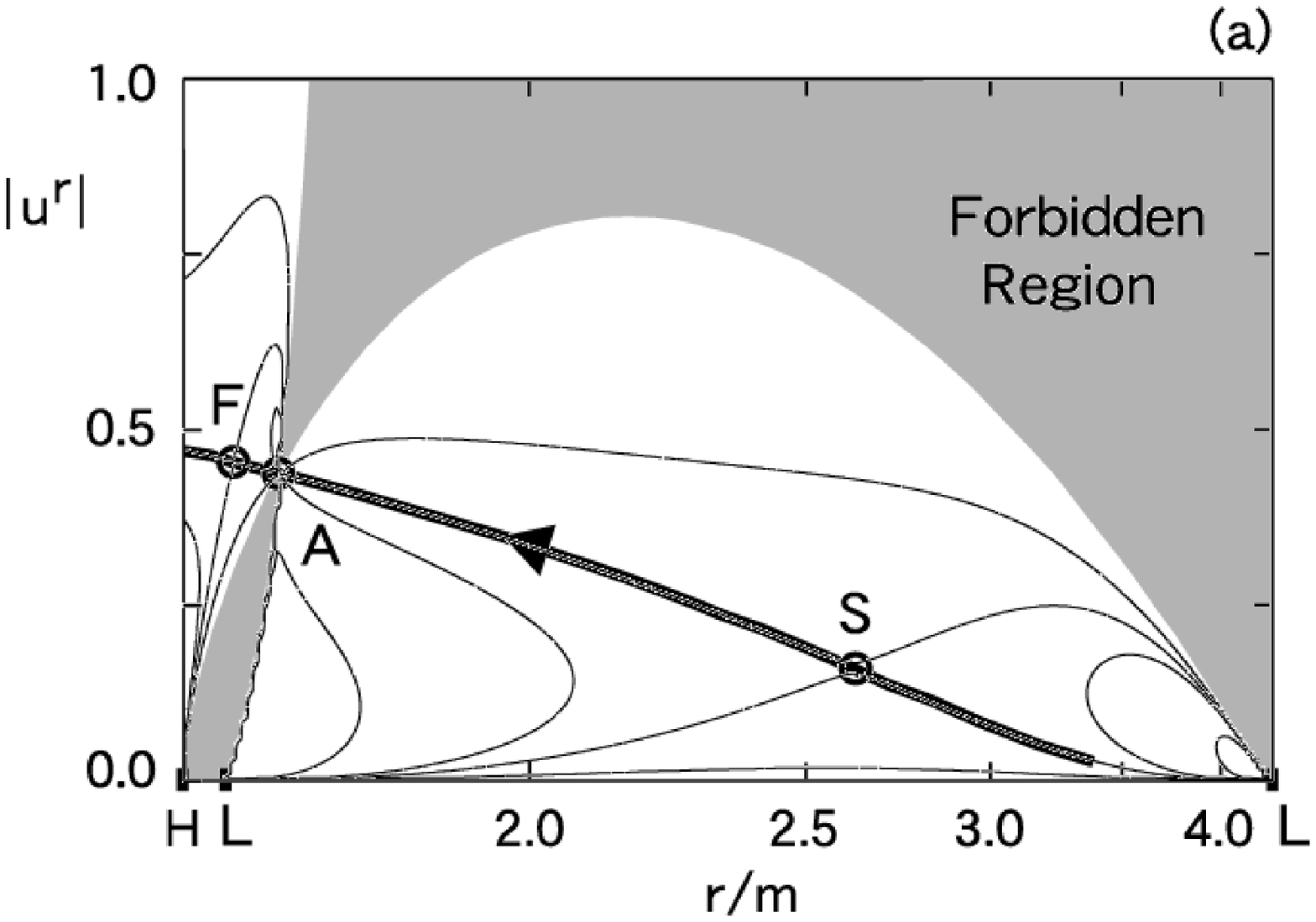} 
   \plotone{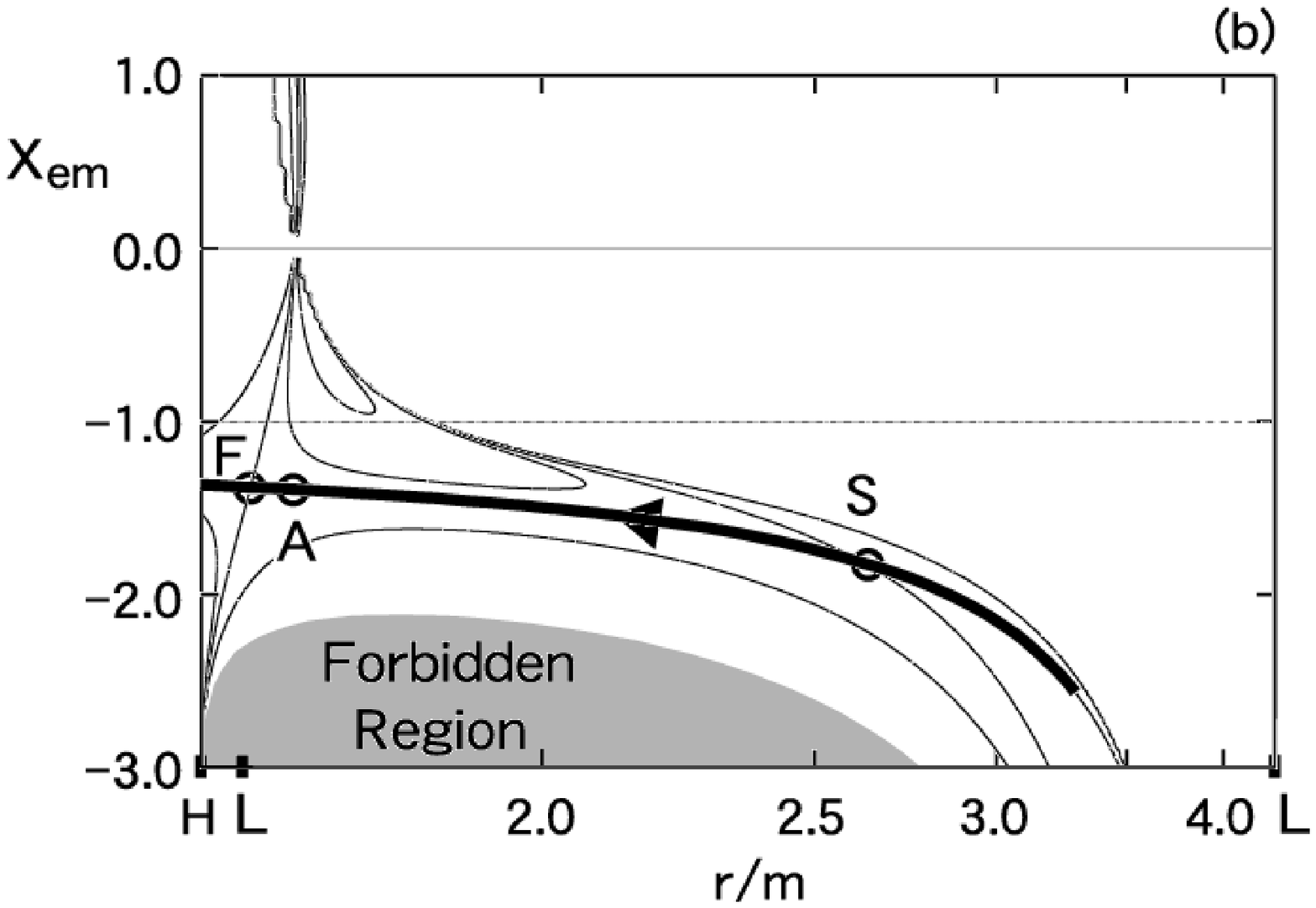} 
\caption{ 
  An example of a negative energy trans-fast MHD accretion 
  solution (thick curves);  with (a) radial 4-velocity and 
  (b) the ratio of the electromagnetic energy to the total energy.
  The flow parameters are $a=0.8m$, $\Omega_F = 0.6\Omega_{\rm max}$, 
  $\delta=0.0$, $\Gamma=4/3$, $\zeta_{\rm F}=0.1$, $x_{\rm A}=0.95$ 
  ($\Omega_F \tilde L=5.1710$) and $r_{\rm F}=1.647934m$, 
  which gives $E_{\rm F}/\mu_{\rm c}= -0.15855$ and $\hat\eta=0.0297$. 
        } 
\label{fig:MHDaccD}
\end{figure} 


\begin{thebibliography}{}
\bibitem[Abramowicz(1981)]{Abramowicz81} 
         Abramowicz, M. A., and Zurek, W. H. 1981, \apj, 246, 314
\bibitem[Bekenstein(1978)]{Bekenstein78}
         Bekenstein, J. D., \& Oron, E. 1978, \prd, 18, 1809 
\bibitem[Beskin(1997)]{beskin97}
         Beskin, V. S. 1997, Physics-Uspekhi, 40, 659
\bibitem[Blandford \& Znajek(1977)]{BZ77}
         Blandford, R. D., \& Znajek, R. L. 1977, \mnras, 179, 433
\bibitem[Camenzind(1986a)]{Camenzind86a}
         Camenzind, M. 1986a, \aap, 156, 137
\bibitem[Camenzind(1986b)]{Camenzind86b}
         Camenzind, M. 1986b, \aap, 162, 32
\bibitem[Camenzind(1987)]{Camenzind87}
         Camenzind, M. 1987, \aap, 184, 341 
\bibitem[Camenzind(1989)]{Camenzind89}
         Camenzind, M. 1989, in Accretion Disks and Magnetic Fields 
         in Astrophysics, ed. G. Belvedere (Dordrecht: Kluwer), 129 
\bibitem[Chakrabarti(1990)]{Chakrabarti90}
         Chakrabarti, S. K., 1990, Theory of Transonic Astrophysical 
         Flows (Singapore: World Scientific)
\bibitem[Heyvaerts \& Norman(1989)]{Heyvaerts-Norman89}
         Heyvaerts, J., \& Norman, C. 1989, \apj, 347, 1055
\bibitem[Hirotani et al.(1992)]{Hirotani-TNT92}
         Hirotani, K., Takahashi, M., Nitta, S., \& Tomimatsu, A. 
         1992, \apj, 386, 455
\bibitem[Kennel, Fujimura \& Okamoto(1983)]{Kennel83}
         Kennel, C., Fujimura, F., \& Okamoto, I. 1983, Geophys. Ap. 
         Fluid Dyn., 26, 147
\bibitem[Lu(1986)]{Lu86}
         Lu, J.F. 1986, Gen.Rel.Grav., 18, 45
\bibitem[Nitta, Takahashi, \& Tomimatsu(1991)]{Nitta-TT91}
          Nitta, S., Takahashi, M., \& Tomimatsu, A. 1991, 
          \prd, 44, 2295  
\bibitem[Phinney(1983)]{Phinney83}
          Phinney, E. S., 1983, Ph.D. thesis, Univ. Cambridge 
\bibitem[Punsly(1990)]{Punsly90}
         Punsly, B. 1990, \apj, 354, 583 
\bibitem[Punsly(2001)]{Punsly01}
         Punsly, B. 2001, Black Hole Gravitohydromagnetics (Springer)
\bibitem[Takahashi et al.(1990)]{Takahashi-NTT90}
         Takahashi, M., Nitta, S., Tatematsu, Y., \& Tomimatsu, A. 
         1990, \apj, 363, 206 (Paper~I)
\bibitem[Takahashi(1994)]{Takahashi94}
         Takahashi, M. 1994, in Proceedings of the Seventh Marcel 
         Grossmann Meeting on General Relativity, ed. R. T. Jantzen,  
         \& G. M. Keiser (Singapore: World Scientific), 1298 
\bibitem[Takahashi \& Shibata(1998)]{Takahashi-Shibata98}
         Takahashi, M., \& Shibata, S. 1998, \pasj, 50, 271
\bibitem[Takahashi(2000)]{Takahashi00}
         Takahashi, M. 2000, in Proceedings of the 19th Texas 
         Symposium on Relativistic Astrophysics and Cosmology, 
         ed. E.~Aubourg, T.~Montmerle, L.~Paul \& P.~Peter  
         (North-Holland, Amsterdam) CD-ROM~01/27.
\bibitem[Thorne, Price, \& Macdonald(1986)]{Membrane86}
         Thorne, K. S., Price, R. H., \& Macdonald, D. A., ed. 1986, 
         Black Holes: The Membrane Paradigm 
         (New Haven: Yale Univ. Press)  
\bibitem[Tomimatsu \& Takahashi(2001)]{TT2001}
         Tomimatsu, A., \& Takahashi, M. 2001, \apj, 552, 710 
\bibitem[Weber \& Davis(1967)]{Weber-Davis67}
         Weber, E. J., \& Davis, L., Jr. 1967, \apj, 148, 217 
\bibitem[Znajek(1977)]{Znajek77}
         Znajek, R. L. 1977, \mnras, 179, 457
\end{thebibliography}
\end{document}